\documentclass[preprint2]{proto}
\usepackage{times}
\usepackage{enumitem}
\pdfoutput=1

\newcommand{\refs}{\par\noindent\hangindent=1pc\hangafter=1}
\newcommand{\tnt}[2]{{\tablenotetext{#1}{#2}}}

\newcommand{\eg}{{\it e.g.}}
\newcommand{\ie}{{\it i.e.}}
\newcommand{\etc}{{\it etc.}}
\newcommand{\etal}{{\it et~al.}}

\newcommand{\oforder}{\mathcal{O}}

\newcommand{\au}{\,\mathrm{au}}
\newcommand{\km}{\,\mathrm{km}}

\newcommand{\meter}{\,\mathrm{m}}
\newcommand{\cm}{\,\mathrm{cm}}

\newcommand{\um}{\,\mu \mathrm{m}}

\newcommand{\yr}{\,\mathrm{yr}}

\newcommand{\second}{\,\mathrm{s}}

\newcommand{\Hz}{\mathrm{Hz}}

\newcommand{\PStwo}{\protect \hbox {Pan-STARRS2}}

\newcommand{\mycitet}[2]{{\it{#1}}~({#2})}
\newcommand{\mycitep}[2]{{\it{#1}},~{#2}}

\begin{document}

\title{\textbf{\LARGE Surveys, Astrometric Follow-up \& Population Statistics}}

\author {\textbf{\large Robert Jedicke}}
\affil{\small\em University of Hawaii}

\author {\textbf{\large Mikael Granvik}}
\affil{\small\em University of Helsinki}

\author {\textbf{\large Marco Micheli}}
\affil{\small\em ESA NEO Coordination Centre}

\author {\textbf{\large Eileen Ryan}}
            \affil{\small\em Magdalena Ridge Observatory}

\author {\textbf{\large Timothy Spahr}}
\affil{\small\em Minor Planet Center}

\author {\textbf{\large Donald K. Yeomans}}
\affil{\small\em Jet Propulsion Laboratory}

\begin{abstract}
\begin{list}{ } {\rightmargin 0.5in} {\leftmargin -1.5in}
\baselineskip = 11pt
\parindent=1pc {\small 
Asteroid surveys are the backbone of asteroid science, and with this
in mind we begin with a broad review of the impact of asteroid surveys
on our field.  We then provide a brief history of asteroid discoveries
so as to place contemporary and future surveys in perspective.
Surveys in the United States have discovered the vast majority of the
asteroids and this dominance has been consolidated since the
publication of Asteroids III.  Our descriptions of the asteroid
surveys that have been operational since that time are focussed upon
those that have contributed the vast majority of asteroid observations
and discoveries. We also provide some insight into upcoming
next-generation surveys that are sure to alter our understanding of
the small bodies in the inner solar system and provide evidence to
untangle their complicated dynamical and physical histories.  The
Minor Planet Center, the nerve center of the asteroid discovery
effort, has improved its operations significantly in the past decade
so that it can manage the increasing discovery rate, and ensure that
it is well-placed to handle the data rates expected in the next
decade.  We also consider the difficulties associated with astrometric follow-up of newly identified objects.  It seems clear that both of these
efforts must operate in new modes in order to keep pace with expected
discovery rates of next-generation ground- and space-based surveys.
\\~\\~\\~}
\end{list}
\end{abstract}

\section{\textbf{INTRODUCTION}}

Without asteroid surveys there would be no asteroid science.  The
cumulative efforts of over 200 years of asteroid surveying has
resulted in the discovery of over half a million asteroids in the
inner solar system that range from just a tenth of an astronomical
unit from the Sun to beyond Jupiter's orbit.  The surveys have
identified asteroids that are the targets of spacecraft missions, that
are the remnants of larger asteroids that were catastrophically
disrupted long ago, and that allow us to untangle the complicated
processes that formed our solar system billions of years ago.  The
surveys' capabilities have improved over the decades as
they took advantage of every new available technology to push their
performance in area coverage, limiting magnitude, and data rates.
Their efforts have enabled our community to advance our understanding
of the past, current and future interactions of both the small and
large objects in our solar system.  This chapter provides a historical
perspective on asteroid surveys, then focuses on their current
capabilities and discoveries since Asteroids III, discusses the
importance of targeted astrometric follow-up for critical objects, and
concludes with a speculative forecast on how the next decade of
asteroid surveying will unfold.

The benefits of a database containing a large number of asteroid orbit
elements and basic physical properties have mostly been achieved over
the past couple decades as a result of the NASA-funded near-Earth
object (NEO) surveys.  They attempt to optimize their surveying for
the discovery of unknown NEOs but, in the process, discover and
recover known asteroids throughout the solar system.  While it seems
obvious that `more is better' it is not necessarily straightforward to
justify the argument for different asteroid populations --- how many
asteroids are necessary for science?  Is a complete survey required or
is a statistical sample sufficient?

The first decade of intensive asteroid surveying at the end of the
last century was motivated by NASA's goal of detecting NEOs that are
larger than $1\km$ in diameter.  The impact on Earth of one of those
objects is expected to have global consequences and the set of about
1,000 NEOs of that size or larger was thought to incorporate 90\% of
the impact risk (\eg\ \mycitep{Morrison}{1992};
\mycitep{Harris}{2008}).  Thus, the surveys had a relatively
well-defined goal that was motivated by planetary defense rather than
science: identify most of the largest hazardous asteroids that could
threaten Earth.  Progress towards achieving that goal could be
measured against the derived population size or by the rediscovery
rate (\eg\ \mycitep{Jedicke \etal}{2003}; \mycitep{Harris}{2007}).  It
is possible that there remain undiscovered large objects on an impact
trajectory with Earth, but the probability is small.

The last decade of asteroid discovery, roughly since Asteroids III,
has mostly retired the risk of an unanticipated impact of a globally
devastating asteroid (\eg\ \mycitep{Harris}{2008}; \mycitep{Mainzer
  \etal}{2011b}). Having achieved that goal, the surveys are now
focussing upon detecting even smaller asteroids with the goal of
discovering $>90$\% of the potentially hazardous objects (PHOs) larger
than $140\meter$ diameter because they 1) represent the lower limit of
those that can cause serious regional ground destruction and 2)
contribute roughly 90\% of the residual hazard to the Earth from
unknown impactors.  The discovery
rate for objects of less than $140\meter$ diameter is currently about 400 per year so that
Asteroids VII could be published before the goal is reached unless new
technologies are brought to bear in the coming decades.

Thus, for PHOs, there are well-defined and clearly-motivated practical
goals for discovering a specific number of objects but the situation
is not so clear for other asteroid populations in the inner solar
system.  How many main belt asteroids are necessary?  To what absolute
magnitude should we strive to be complete for Jupiter's Trojan
asteroids?  Without the NEO surveys driving the discovery of asteroids
to fainter apparent magnitudes and smaller sizes we would not have
discovered, for example, extremely young asteroid families that can be
traced back in time to their collision origin
(\eg\ \mycitep{Nesvorn{\'y} \etal}{2006}; Nesvorn{\'y} \etal, this
volume), widely-separated asteroid pairs on nearly identical orbits
that point to YORP spin-up and tidal disruptions of large asteroids
(\eg\ \mycitep{Vokrouhlick{\'y} and Nesvorn{\'y}}{2008}; Walsh \etal,
this volume; Vokrouhlick{\'y} \etal, this volume), main belt comets
that provide evidence of a water reservoir that may have allowed life
to thrive on Earth (\eg\ \mycitep{Hsieh and Jewitt}{2006};
\mycitep{Jewitt}{2012}; Jewitt \etal, this volume), and contemporary
catastrophic disruptions of main belt asteroids that suggest that YORP
driven rotational spin-up might be the dominant cause of their breakup
(\eg\ \mycitep{Jewitt \etal}{2010}; \mycitep{Denneau \etal}{2015}).
In addition, we would not have been able to measure the ages of main
belt asteroid families through fitting the characteristic
Yarkovsky/YORP-induced `V' shape of the families' absolute magnitude
vs. proper semi-major axes distributions
(\eg\ \mycitep{Vokrouhlick{\'y} \etal}{2006}; Vokrouhlick{\'y} \etal,
this volume) to constrain the dynamical and collisional evolution of
the main belt since its formation (\eg\ \mycitep{O'Brien and
  Greenberg}{2005}; Morbidelli \etal, this volume; Bottke \etal, this
volume).

The surveys' capabilities are now so synoptic that most slow moving
objects brighter than about $V=21$ are routinely detected multiple
times in a lunation, and will be re-detected regularly in the future,
so that they require no specific allocation of resources for targeted
follow-up observations.  Indeed 80\% of the known main belt asteroids
with $H<16.5$ are now numbered, meaning that their orbits are good
enough to predict their future ephemerides to with $2\arcsec$ for at
least the next decade.  On the other hand, rapidly moving, nearby NEOs
that may pose an Earth impact hazard in the future usually require
rapid follow-up to measure their astrometric positions (\mycitep{Ticha
  \etal}{2002}) over as long an arc as possible during their discovery
opposition to 1) accurately assess their Earth impact hazard
(\eg\ \mycitep{Milani \etal}{2005}; Harris \etal, this volume) and 2)
increase the probability that detections of the object in future
apparitions can be associated with the discovery apparition
observations (\eg\ \mycitep{Milani \etal}{2012}; Farnocchia \etal,
this volume).

We think that continued asteroid survey efforts in the next
decades are justified in order to expand upon the rich
asteroid science yield of the past decades, to provide exciting new
discoveries that will generate unexpected insights into our solar system's formation and continuing evolution, and to further reduce the NEO impact hazard risk.  The remainder of this chapter provides an introduction to
the history of asteroid surveys before Asteroids III, the continuing
survey improvements and their current status as of publication of
Asteroids IV, the importance and state-of-the-art of follow-up
efforts, and our perspective on upcoming surveys and technologies as
we look forward to Asteroids V.

\bigskip
\refstepcounter{section}
\noindent{\textbf{ \thesection. A BRIEF HISTORY OF ASTEROID DISCOVERY (PRIOR TO ASTEROIDS III)}}
\bigskip

The first asteroid to be discovered, (1)~Ceres, was visually
identified by Father Giuseppe Piazzi, director of the Palermo
Observatory, on the morning of 1801 January 1, the first day of the
nineteenth century.  Piazzi was identifying and correcting the
positions of stars in an existing catalog when he noticed that one
`star' in Taurus shifted position from night to night.  He followed
the object until 1801 February 11 but his discovery was not announced
until the summer of that year.  It was recovered on 1801 December 7
only after Karl Gauss, with characteristic genius, provided an
elliptic orbit for (1)~Ceres that allowed an accurate-enough ephemeris
prediction.  Three more asteroids were discovered shortly thereafter:
Heinrich Olbers identified the second and fourth asteroids, (2)~Pallas
and (4)~Vesta, in 1802 and 1807 respectively while Karl Harding
discovered (3) Juno in 1804.  It was another 41 years until
(5)~Astraea was discovered in 1845, and the discovery rate would
remain less than about one per month till about the time of the advent
of astrophotography.  There were only ten known asteroids by
mid-century and 447 by 1900 (see
Figure~\ref{fig.MBO+NEO-Discoveries-per-5years}).

The era of photographic asteroid discovery began in 1891 when Max Wolf
compared the images on three successive photographic plates to detect
the motion of (323)~Brucia, the first of Wolf's 321 asteroid
discoveries.  Trailed detections of asteroids on single long-exposure
photographic plates were also used to identify asteroids and Gustav
Witt discovered the trailed image of the first near-Earth object,
(433)~Eros, at the Urania Observatory in Berlin in 1898.

The pioneering photographic asteroid surveys in the 1970s and early
1980s marked the beginning of the modern NEO discovery era (see
Figure~\ref{fig.MBO+NEO-Discoveries-per-5years}).  Gene Shoemaker and
Eleanor Helin began using the 18-inch Palomar Schmidt telescope in
southern California, USA, for finding NEOs in 1973 and were joined by
Carolyn Shoemaker in the early 1980s (\mycitep{Helin and
  Shoemaker}{1979}).  Their early photographic surveys identified
asteroids using manually operated blink comparators and
stereomicroscopes that enabled visual comparison of images of the same
portion of sky taken several minutes apart.  The vast majority of the
objects in the images were stationary stars and galaxies but a moving
NEO would be in a slightly different position on each photograph so
that it would appear to jump back and forth when each image was
quickly viewed in turn with the blink comparator.  Alternately, the
NEO's image would appear to `float' above the background stars when
two different images were examined at the same time with a
stereomicroscope.  (For more details concerning these pioneering NEO
search efforts see; \eg\ \mycitep{Cunningham}{1988}; \mycitep{Stokes
  \etal}{2002}; \mycitep{Yeomans}{2013})

The surveys entered the modern era in 1984 when the Spacewatch
telescope (see \S\ref{ss.Spacewatch}) became the first survey to
employ a camera with a charge-coupled device (CCD) focal plane (see
Figure~\ref{fig.MBO+NEO-Discoveries-per-5years}). Their first
$320\times512$ pixel CCD detector was replaced in 1989 with a large
format $2048\times2048$ CCD that was used for three years until they
obtained a high-efficiency ($\sim70$\%) thinned $2048\times2048$ CCD.
The system was operated for about 23 nights per month with the CCD
read out in a time-efficient `drift-scanning' mode in which the right
ascension axis was stationary so that the star field would drift
through the telescope's field-of-view (FOV) while the CCD detector was
read out at the same rate.  This technique allowed the survey to image
about 200~deg$^2$ each month to a limiting $V$-band magnitude of $\ga21$.  Each scan was repeated
three times with about thirty minutes separation and automated
software identified moving objects in the field
(\mycitep{Rabinowitz}{1991}).

The NEO discovery rate increased dramatically in the late 1990s (see
Figure~\ref{fig.MBO+NEO-Discoveries-per-5years}) when NASA increased
the funding available for NEO surveys, partly in response to impact
awareness generated by the 1994 impact of Comet Shoemaker-Levy~9 with
the planet Jupiter.  The perfect storm of increased funding coupled
with the decreasing cost of CCDs and the availability of
frame-transfer CCDs enabled the rise of the LINEAR (\S\ref{ss.LINEAR})
and Catalina (\S\ref{ss.CSS}) NEO surveys that dominated the modern
survey era for about two decades leading up to and through Asteroids
III.

(\mycitet{Stokes \etal}{2002} provides more details of the history and
state of asteroid surveys leading up to Asteroids III.)

\bigskip
\refstepcounter{section}
\centerline{\textbf{ \thesection. ASTEROID SURVEYS (since Asteroids III)}}
\label{s.AsteroidSurveys}
\bigskip

All contemporary asteroid surveys use extremely efficient CCDs to
record digital images of the sky.  While CCD detectors are far more
sensitive and accurate than film their application to asteroid
discovery is similar.  Three or more CCD images are taken of the same
region of the sky with successive images separated in time by about 30
minutes.  The images are then compared in software to identify
detections that have systematically moved to different positions from
one image to the next.  The rate of motion of the detections from one
image to the next, the direction they appear to be traveling, and
their apparent brightness are helpful in identifying interesting
objects and can provide first-order estimates of an object's distance
from Earth, its size and general orbital characteristics.  For
example, an object that appears to be moving very rapidly from one
image to the next ($>1\,$deg/day) is almost certainly an NEO.  New NEO
discoveries are usually still verified with the human eye even though
sophisticated and automated software analyses of the CCD images have
replaced manual identification of moving objects.

In 1998, NASA established the goal of discovering 90\% of NEOs larger
than one kilometer in diameter and in 2005 Congress extended that goal
to include 90\% of the NEOs larger than 140\,meters (George
E. Brown,~Jr. Near-Earth Object Survey Act).  There are thought to be
about 1,000 near-Earth objects larger than one kilometer diameter and
roughly 26,000 larger than 140 meters (\eg\ \mycitet{Granvik
  \etal}{2015}; Harris \etal, this volume).  The desire to meet the
NEO goals has enabled the funding and driven the success of the
asteroid surveys over nearly the past two decades.

Asteroid surveys that search the largest volume of sky each month will
discover the most NEOs (\mycitep{Bowell and Muinonen}{1994}), all
other things being equal.  How much sky each telescope surveys depends
upon several factors including the number of clear nights available
for observing, the telescope aperture and FOV, and the sensitivity and
efficiency of the CCD detector.  That being said, not all regions of
the sky are equally productive for discovering new NEOs
(\eg\ \mycitep{Bowell and Muinonen}{1994}; \mycitep{Chesley and
  Spahr}{2004}; \mycitep{Vere{\v s} \etal}{2009}).  Once the most
effective regions are fully surveyed it is important to extend the
search to greater distances from Earth or, in other words, to fainter
limiting magnitudes.

Space-based asteroid surveys have advantages over ground-based
observing in that they are not hindered by weather, they can search
continuously, can observe in the infrared where asteroids are brighter
and there are fewer background sources, and space telescopes can
observe NEOs when they are much closer to the Sun (Mainzer \etal, this
volume). A case in point is the $\sim18\meter$ diameter asteroid that
injured more than 1,500 people when it airburst over Chelyabinsk,
Russia, on 2013 February 15 that was not detected by ground-based
surveys since it came from a sunward direction (\eg\ \mycitep{Brown
  \etal}{2013}). Of course, space-based surveys are expensive, risky,
can be data-rate limited due to the availability and limitations of
the downlink from the spacecraft to the ground, and they are not
repairable in the event of a major failure.

The asteroid surveys have discovered about 90\% of the NEOs larger
than $1\km$ diameter since NASA's initiation of its NEO Observations
(NEOO) program in 1998 (\mycitep{Mainzer \etal}{2011b}), and a good
fraction of those larger than $140\meter$.  Progress toward meeting
the goals can be monitored on the NEO discovery statistics page at
{\tt neo.jpl.nasa.gov/stats/}.  The vast majority of those NEO
discoveries were made by NASA-supported ground-based telescopic
surveys (see table~\ref{tab.Top10DiscoverySurveys}) and in the
following sub-sections we briefly describe the major surveys that were
or are operational since Asteroids III in roughly chronological order
by start date (JPL maintains a list of NASA-supported NEO survey
programs at {\tt http://neo.jpl.nasa.gov/programs/}).

\bigskip
\noindent
\refstepcounter{subsection}
\textbf{\thesubsection\ Spacewatch (1983-present)}
\label{ss.Spacewatch}
\smallskip

The Spacewatch team was a pioneer in digital detection of NEOs
(\eg\ \mycitep{Rabinowitz}{1991}).  They discovered the first NEO on
digital images in 1989 and reported the first automated (software)
discovery in 1990.  They led the effort to automate the discovery and
follow-up of NEOs, culminating in the modernization of their
$0.9\meter$ and $1.8\meter$ telescopes that can both be operated from
a single control room by a single observer.

They began using their custom-built $1.8\meter$ aperture telescope for
NEO surveying and follow-up in 2002.  Later that same year their
$0.9\meter$ was instrumented with a large-scale mosaic camera
consisting of four $4608\times2048$ CCDs to take advantage of its new
optical system that provides a 2.9\,deg$^2$ FOV. The $0.9\meter$
telescope's new optical configuration required that it operate in the
conventional `stare' mode whereas the $1.8\meter$ telescope continued
to be operated in the `drift-scan' mode until 2011.  From 2005 through
2008 Spacewatch gradually shifted its emphasis from NEO surveying to
follow-up as other surveys began to dominate NEO discovery.  Finally,
the imaging detector on the $1.8\meter$ telescope was replaced in 2011
October with a stare-mode CCD with finer pixel resolution, faster
readout, and flatter focus.  The new camera allowed Spacewatch to
increase their NEO follow-up rate by more than 50\% while at the same
time halving the astrometric residuals.  The Spacewatch $1.8\meter$
telescope is currently among the world's leaders in faint-object
follow-up, especially in terms of critical follow-up of the most
challenging faint objects.

\bigskip
\noindent
\refstepcounter{subsection}
\textbf{\thesubsection\ Near-Earth Asteroid Tracking (NEAT, 1995-2007)}
\smallskip
 
Beginning in 1995 the JPL Near-Earth Asteroid Tracking program (NEAT,
\mycitep{Pravdo \etal}{1999}) operated survey telescopes on the summit
of Haleakala, HI, USA, in cooperation with the Air Force.  The GEODSS
$1\meter$ telescope was equipped with a 4k$\times$4k CCD and initially
utilized 12 nights per month centered on new moon (1995 December to
1996 December) subsequently reduced to 6 nights per month (1997
January to 1999 February).  In 2001 February NEAT began using a
modified $1.2\meter$ telescope, part of the U.S. Air Force Space
Surveillance System on Maui, HI, USA.  Nightly pointing lists were
generated at JPL for the telescope control computer in Maui and small
sub-images of candidate NEOs were transferred back to JPL for
inspection by JPL scientists.  In 2001 the NEAT program transitioned
to the $1.2\meter$ Schmidt telescope at Palomar Mountain in southern
California, USA, before ceasing operations in 2007.

\bigskip
\noindent
\refstepcounter{subsection}
\textbf{\thesubsection\ Lowell Observatory NEO Survey (LONEOS, 1993-2008)}
\smallskip

The Lowell Observatory NEO Survey (LONEOS; \mycitep{Bowell
  \etal}{1995}) operated from 1993 through 2008 using the Lowell
Observatory $0.6\meter$ Schmidt telescope at Anderson Mesa near
Flagstaff, AZ, USA.  The relatively small aperture telescope was
competitive because it had a large FOV of about 8\,deg$^2$
instrumented with two 2k$\times$4k cooled CCDs.  LONEOS collected four
45$\second$ exposures of each field that were automatically searched
for moving objects down to $V\sim19.3$.  The LONEOS system's
productivity eventually declined due to competition with larger
aperture survey telescopes so they switched their primary objective to
photometric observations of NEOs and finally ceased operations in
2008.

\bigskip
\noindent
\refstepcounter{subsection}
\textbf{\thesubsection\ Lincoln Near-Earth Asteroid Research program (LINEAR, 1996-2013)}
\label{ss.LINEAR}
\smallskip

The LINEAR survey (\mycitep{Stokes \etal}{2000}) was operated out of
Socorro, NM, USA, by the MIT-Lincoln Laboratory team that was based in
Lexington, MA, USA.  The legacy $1\meter$ LINEAR telescope system is
located at Lincoln Laboratory's Experimental Test Site near Stallion
Range Center on the US Army's White Sand Missile Range in central New
Mexico, USA.  It began operations in early 1996 but was in routine use
from 1998 March through 2013 May.  The survey used two $1\meter$
aperture telescopes originally designed as prototypes for
Earth-orbiting debris tracking.  LINEAR's success was driven by the
application of electro-optical sensor technology, originally developed
for US Air Force Space Surveillance, to the problem of discovering
near-Earth asteroids and comets.  Their frame-transfer CCDs allowed
large areas of sky to be surveyed extremely efficiently because the
camera required effectively zero readout time.  Furthermore, the
rapid-readout CCDs and access to fast-computing resources allowed them
to apply an advanced image processing technique involving generating a
median of 5 images of the same field and subtracting the median from
the sum of the same 5 images to generate difference images that
contained only transients and relatively few artifacts.  The
differenced images were then searched for transient detections moving
in roughly straight lines (tracklets).  During the late 1990s and
early 2000s they could survey several thousand square degrees per
night to $V\sim19$ enabling the program to singlehandedly discover
more than a third of all NEOs one kilometer in diameter or larger.
Lincoln Laboratory discontinued use of the $1\meter$ system in 2013
while it transitions to the use of the $3.5\meter$ Space Surveillance
Telescope (SST, see \S\ref{ss.NewFacilities}).

\bigskip
\noindent
\refstepcounter{subsection}
\textbf{\thesubsection\ Catalina Sky Survey (CSS, 1992-present)}
\label{ss.CSS}
\smallskip

The University of Arizona's Catalina Sky Survey (CSS, \mycitep{Larson
  \etal}{1998}) has been the primary NEO discovery system over much of
the time since Asteroids III and has done so while remaining
cost-competitive.  Their success can be traced to remaining focussed
on their primary objective of discovering NEOs; employing dedicated,
professional, skillful observers; and by standardizing equipment and
software across their different sites.  For instance, all the cameras
at their three sites are identical, thinned, 4k$\times$4k
back-illuminated detectors packaged by Spectral Instruments, Inc., of
Tucson, AZ, USA, that are cooled with a closed-cycle cryocooler.  The
use of identical cameras has minimized hardware and software
development, maintenance and operations costs.  Their dedication to
continual improvement, comprehensive sky coverage, human vetting of
candidate discoveries, and on-site recovery capability have further
contributed to their success over the last decade.

The CSS was preceded by the Bigelow Sky Survey (BSS) that photographically searched for new asteroids and comets between 1992 and 1996 (\mycitep{Spahr \etal}{1993}; \mycitep{Spahr \etal}{1996}).  The experience gained in operating the BSS informed and motivated modifications to the telescope, camera and software beginning in 1997 so that the CSS's CCD survey began producing observations in 1998. 

The CSS has used three different telescopes in the past ten years and
in 2013 alone discovered 603 NEOs, easily making it the top survey
program that year.  Their two primary telescopes are the Catalina
Schmidt (703) and the Mt. Lemmon telescopes (G96) but their third
site, the Siding Spring Schmidt (E12) in Australia, has delivered
enough NEOs to place it in the top 6 NEO discovery sites in each of
the past 10 years.  About 20\% of CSS observing time is devoted to
post-discovery follow-up observations (see \S\ref{s.follow-up}).

The Catalina Schmidt (703) in the mountains north of Tucson, AZ, USA,
was upgraded to a $f$/1.8, $0.7\meter$ telescope from 2003 to 2004
which increased their sensitivity to $V\sim20.0$.  Its 8.2\,deg$^2$
FOV enabled the site to survey 1,000 to 1,500$\,\deg^2$ per night and
vaulted the CSS program into the lead in terms of annual discoveries.

In 2004-2005 the CSS team procured a 4k$\times$4k CCD and then built
and installed a prime-focus camera that delivers a 1.2\,deg$^2$ FOV on
their Mount Lemmon $1.5\meter$ telescope (G96).  Its limiting
magnitude reaches $V\sim21.5$ under good conditions and they can
survey $\sim200$\,deg$^2$ on an average night.  This survey excels at
finding very small objects that are very close to Earth
(\mycitep{Jedicke \etal}{2014}) and demonstrates the need to shift to
larger optical systems in order to complete the inventory of smaller,
yet still threatening, NEOs.  \eg\ The CSS identified the only two
Earth impactors that were discovered in advance of impact
(2008~TC$_3$, \mycitep{Kowalski \etal}{2008}; 2014~AA,
\mycitep{Kowalski \etal}{2014}).

The $0.5\meter$ Uppsala Schmidt telescope (E12) at Siding Spring,
Australia, was operated by the CSS for several years when it was the
only NEO survey in the southern hemisphere.  However, in 2013 the
Australian dollar became much stronger relative to the
U.S. dollar-based funding from the NASA NEOO program compared to when
the survey was established.  The combination of the exchange rate
shift and the telescope's modest aperture and FOV reduced the
cost-effectiveness of the survey compared to the CSS's northern
hemisphere assets and the decision was made to end support for this
facility.

\bigskip
\noindent
\refstepcounter{subsection}
\textbf{\thesubsection\ Panoramic Survey Telescope and Rapid Response System (Pan-STARRS, 2010-present)}
\label{ss.PanSTARRS}
\smallskip

The $1.8\meter$ Pan-STARRS1 (PS1; \mycitep{Wainscoat \etal}{2013})
telescope on Haleakala, HI, USA, was developed by the University of
Hawaii's Institute for Astronomy and operated by the PS1 Science
Consortium (PS1SC) until early 2014.  It was the prototype telescope
for the 4-telescope Pan-STARRS system (\mycitep{Kaiser \etal}{2002})
that was supposed to be operational late in the last decade but may
never be completed. The PS1 CCD camera was the largest in the world
when it was originally built, with a focal plane consisting of an
almost complete $8\times8$ array of CCDs (the four corner CCDs were
left out of the focal plane) with a total of about 1.4~gigapixels. The
large focal plane combined with the system's $f/4$ optics yields a
$\sim7\,\deg^2$ FOV that has moved PS1 to be the leading PHO discovery
site beginning in calendar year 2012.  In 2014 it will discover about
$4\times$ more PHOs than the second most successful site.
 
The first PS1 NEO discoveries were recorded in the second half of 2010
but only about 5\% of the observing time was devoted to NEO
discoveries at that time.  The NEO survey time fraction was increased
to 11\% beginning in 2012 November. In addition, 56\% of the observing
time was used for a `$3\pi$' sky survey in three filters that was also
executed in a manner that led to the discovery of NEOs.  The time
devoted to the NEO search was increased to 100\% beginning in 2014
April with funding provided by NASA's NEOO program after the end of
the PS1 Science Consortium.

PS1's strength lies in having the faintest magnitude limit of any
active NEO survey, reaching $V\sim22$ under good conditions.  This
allows the team to discover NEOs that are too faint to be detected by
the other systems.  Their excellent site in the middle of the Pacific
Ocean allows PS1 to survey the `sweet spots' near the Sun where the
sky-plane PHO density is highest if the system can reach fainter than
$V\sim21$ (\mycitep{Chesley and Spahr}{2004}, \mycitep{Vere{\v s}
  \etal}{2009}).

PS1 developed a sophisticated image processing pipeline (IPP,
\mycitep{Magnier}{2006}) that feeds transient detections to the Moving
Object Processing System (MOPS, \mycitep{Denneau \etal}{2013}), that
uses kd-trees to quickly link transient detections into tracklets
(\mycitep{Kubica \etal}{2007}).  The final observations are extremely
accurate with astrometry good to $\la0.1\arcsec$
(\mycitep{Milani}{2012}) that allows the MPC to tightly constrain NEO
orbital solutions and uncertainty maps to facilitate the recovery of
PS1 NEO candidates.

\bigskip
\noindent
\refstepcounter{subsection}
\textbf{\thesubsection\ NEO Wide-field Infrared Survey Explorer (NEOWISE, 2010-present)}
\smallskip

The NEOWISE program (\eg\ \mycitep{Mainzer
  \etal}{2011a}; Mainzer \etal, this volume) observes and discovers NEOs and other asteroids in
the near-IR with the $0.4\meter$ telescope aboard the Wide-field
Infrared Survey Explorer (WISE) spacecraft (\eg\ \mycitep{Cutri \etal}{2012}; \mycitep{Wright \etal}{2010}).  The spacecraft was
launched on 2009 December 14 into a `sun-synchronous' polar orbit
around Earth (the spacecraft's orbital plane is always roughly
perpendicular to the Earth-Sun line) and as it orbited it continuously surveyed a $47\arcmin$ wide strip of sky in the opposite direction from Earth.  WISE operated for ten months in
2010 performing an all-sky astronomical survey in four bands centered
at 3.4, 4.6, 12 and 22$\um$.  When the hydrogen coolant for the two
longest wavelength detectors was exhausted, a secondary post-cryogenic
mission continued for four more months using just the two shorter
wavelength detectors, and then the spacecraft was decommissioned and
hibernated on 2011 February 17.  After more than two and a half years
the spacecraft was awakened from hibernation and NEOWISE was
reactivated in 2013 September for a planned 3~year observing period
using just the $3.4\um$ and $4.6\um$ passbands
(\mycitep{Mainzer}{2014}).  Its first NEO discovery in the new
operational phase, 2013~YP$_{139}$, occurred on 2013 December 29.

NEOWISE's Sun-synchronous survey mode imposes a time limit of about 36
hours during which
images at a specific location can be acquired.  Hence, the program requires ground-based NEO candidate
follow-up by a dedicated network of amateur and professional
astronomers to secure their orbits (see
\S\ref{s.follow-up}).  NEOWISE also detected tens of thousands of known
asteroids in the IR that allowed diameter measurements of thousands of
objects thus enabling a range of studies of the origins and evolution
of the small bodies in our solar system (\eg\ NEO, \mycitep{Mainzer
  \etal}{2011b}; MBO, \mycitep{Masiero \etal}{2014}; Jupiter Trojans,
\mycitep{Grav \etal}{2012}; and Mainzer \etal, this volume).

\bigskip
\noindent
\refstepcounter{subsection}
\textbf{\thesubsection\ Other Contributions}
\smallskip

Amateur astronomers have historically played a role in advancing the
astronomical fields but their contribution to NEO discovery has been
limited due to their access to telescopes of relatively modest
apertures.  From 1998 through 2013 only about 1.8\% of the 10,044
discovered NEOs were found by the amateur community.  The most
productive team was the Las Sagra Survey (LSS) operated in Spain by
J. Nomen, R. Stoss, and others.  They discovered 79 NEOs prior to
being funded professionally by the European Space Agency for space
debris tracking.  During the same time period only 99 other NEOs were
discovered by amateurs from the rest of the world combined.

A small number of asteroids are discovered serendipitously by
professional observers or other astronomical surveys in the course of
their work that is not always primarily associated with asteroids.  For
instance, the International Scientific Optical Network (ISON;
\mycitep{Molotov}{2010}) is designed to identify and track
Earth-orbiting space-debris but it's capabilities naturally serve the
asteroid identification processes illustrated by their discovery of
the spectacular Comet ISON (C/2012 S1).  Similarly, the Palomar
Transient Factory (PTF; \eg\ \mycitep{Polishook \& Ofek}{2011}) is an
all-sky astronomical survey designed to identify transient and
variable objects that has also identified many asteroids and NEOs.

\bigskip
\refstepcounter{section}
\centerline{\textbf{ \thesection. THE MINOR PLANET CENTER (MPC) }}
\label{s.MPC}
\bigskip

All the candidate detections of minor planets identified by the
surveys are delivered to the Minor Planet Center (MPC, {\tt
  http://minorplanetcenter.net}) --- the IAU's official international
repository and distributor of asteroid astrometric observations, minor
planet orbits, and identifications.  The MPC's goal is to identify
and/or link all reported minor planet observations in nearly real
time.  This is a challenging process because the MPC has experienced
a $7\times$ increase in reported observations over the time since
Asteroids III (see Figure~\ref{fig.MBO+NEO-Discoveries-per-5years})
and the time required for orbit determination and linking increases
faster-than-linearly with the number of observations in the database.
Fortunately, Moore's Law and the implementation of improved
operational procedures have allowed the MPC to handle the increased
data and analysis rate.  Today, the MPC can automatically and
seamlessly receive and process a few million observations of a few
hundred thousand minor planets each day from both ground- and
space-based observatories.

The MPC concentrates on expediently processing NEO discoveries and
observations because they are of interest to NASA and the
public. Ephemerides need to be provided promptly so that follow-up
telescope facilities can quickly recover the objects.  The MPC checks
all reported NEO candidate `tracklets' (sets of detections that are
claimed to be of the same object) for identification with known
objects, computes the likelihood that unknown objects are new NEOs,
and posts the best candidates on the NEO Confirmation Page (NEOCP) for
follow-up (the NEOCP is available at {\tt
  http://www.minorplanetcenter.net/iau/NEO/ToConfirm.html}; follow-up
efforts are described in \S\ref{s.follow-up}).  The ESA's NEO Coordination Centre also maintains a parallel and prioritized follow-up list at
{\tt http://neo.ssa.esa.int/web/guest/priority-list}.

The NEOCP was handily validated in 2008 October when the MPC
identified a CSS discovery that would impact Earth only 20 hours after
discovery (\mycitep{Kowalski and Chesley}{2008}). The international
network of observers monitoring the NEOCP quickly responded (see
\S\ref{s.follow-up}) and reported more than 800 astrometric
observations, multiple light curves and even a spectrum, over a
timespan of less than a day. As a result of this coverage JPL
predicted that the object now known as 2008~TC$_3$ would impact in a
small desert area in northern Sudan. This prediction was subsequently
confirmed by visual and satellite reports and with the collection of
meteorite fragments on the ground (\eg\ \mycitep{Jenniskens
  \etal}{2009}; Jenniskens \etal, this volume).

The MPC's database (see table~\ref{tab.MPCHoldings}) mostly contains optical observations of main belt
asteroids even though much of the MPC's time and effort is devoted to
dealing with NEOs.  As of this writing there are almost 400,000 minor
planets classified as `numbered' (most of them in the main belt),
meaning that their orbits are of sufficient accuracy that their
ephemeris uncertainties in the next decade are no more than a few
arc-seconds.  To be numbered an
object usually needs to be observed over a period of at least 4
oppositions with detections on more than 1 night in most of the
apparitions.  This rule-of-thumb applies to main belt objects that
typically have only optical sky-plane positions --- NEOs may have
secure orbits with just two apparitions of data because of the power
of parallax for nearby objects to reduce the uncertainty on the orbit
elements.  Similarly, the range and range-rate accuracy available with
radar may be combined with optical observations from just a single
apparition to yield good orbits (\eg\ \mycitep{Ostro \etal}{2004}).
In total, the numbered objects have increased by more than a factor of
10 since Asteroids III and about 90\% of those objects were discovered
by just ten surveys or sites (see
table~\ref{tab.Top10DiscoverySurveys}).  Many of the objects that are
now numbered were discovered more than a decade ago, before Asteroids
III, because it requires about 10 years before a typical main belt
object has been detected in enough apparitions to allow it to be
numbered.  Thus, the tremendous number of new discoveries by
contemporary NEO surveys (see \S\ref{s.AsteroidSurveys}) will only be
numbered as this decade progresses towards the publication of
Asteroids V.

The unfortunate reality of the difference between the capabilities of
professional and amateur facilities is that only 0.6\% of the numbered
objects in the top 10 list were contributed by amateur observatories.
Eight of the top ten most productive programs are, or were, funded
directly by NASA specifically for discovery or follow-up of
NEOs. Thus, it is clear that the discovery of minor planets is almost
uniquely a NASA-funded survey effort based in the United States.
Indeed, fully 80\% of the discoveries were realized at just a handful
of NASA-funded NEO search programs led by the LINEAR program (see
\S\ref{ss.LINEAR}).

The MPC database contains observations and orbits of objects that were
detected in multiple oppositions, a single opposition, and even on
only two nights.  The sum total of all these objects is currently
comparable to the number of numbered objects (see
table~\ref{tab.MPCHoldings}).  The accuracy of the ephemerides for
each successive class of objects decreases dramatically such that
single opposition orbits might be suitable for predicting the location
of the object for only another 1 or 2 years while objects observed on
only 2 nights can typically only be used for about 10 days.

Finally, the MPC maintains a publicly available file of all
single-night detections that currently cannot be linked or identified
with other minor planets.  This pejoratively dubbed (by Brian Marsden
in the mid 1990s) `One Night Stand' (ONS) file now contains over 8
million observations.  While it is likely that many of the
observations in this file represent duplicates, false detections, and
errors, it is also likely that the file includes several hundred
thousand uncatalogued minor planets.  The MPC currently employs a
graduate student (J. Myers, U. of Arizona) who is developing advanced
linking techniques to extract as many objects as possible from the ONS
file but it will always contain a high fraction of unlinkable and
likely false detections.

One challenge at the MPC is linking and processing data sets of very
different observational quality.  For example, the mean astrometric
residual for observations prior to the CCD era (c. 1995) is about
1-2\arcsec\ while residuals from the modern Pan-STARRS program
(\S\ref{ss.PanSTARRS}) are $\sim0.1\arcsec$ --- a factor of 10 to 20
improvement!  Even today, the mean residuals for survey data from the
professional surveys ranges from about 0.05\arcsec to 0.7\arcsec .

\bigskip
\refstepcounter{section}
\centerline{\textbf{ \thesection. ASTROMETRIC FOLLOW-UP,}}
\centerline{\textbf{ RECOVERY AND PRECOVERY}}
\label{s.follow-up}
\bigskip

Modern survey telescopes are specifically designed to have a wide FOV
and rapid readout in order to image as much sky as possible so that
using them for targeted follow-up of specific objects is an
inefficient use of their capabilities.  Instead, the surveys identify
and report likely asteroid tracklets to the MPC on a nearly real-time
or nightly basis and then continue to survey the sky for other
candidates.  They rely on other `follow-up' facilities to obtain more
observations of the objects for verification and to secure their
orbits.  Follow-up usually refers to obtaining astrometric and
photometric observations of an object to secure its orbit and
determine its absolute magnitude but may also include various types of
`physical characterization' including obtaining an object's light
curve to determine its rotation rate and, perhaps, shape, or spectra
to determine its taxonomy (and mineralogy).  This chapter and sub-section will concern itself almost exclusively with astrometric follow-up.

The most challenging
follow-up is for the NEOs.  More distant objects tend to move slowly
so that their ephemeris uncertainties are relatively small and, in any
event, they will be re-detected in the course of normal ongoing survey
operations.  The first detections of a new candidate NEO typically span a temporal
arc of less than an hour and represent just the beginning of the
lengthy and complex discovery process.  First, the candidate must be
confirmed as being a real and unknown object --- it is not uncommon
for reported candidates to be false because of processing errors such
as mis-linking of two different asteroids or combining false
detections in an image into a false tracklet.  Once an object is
established as being real and recoverable, discovery observations must
be followed by an observational campaign to establish the object's
orbit with enough accuracy to enable it to be `recovered' in the
future.  (Recovery is the process of obtaining new observations that
can be attributed to an object.) The extension
of the observed arc is essential for more accurate orbit determination
that leads to a better understanding of the object's dynamics either
because of its scientific interest or because of its possible impact
threat to Earth (\eg\ \mycitep{Milani \etal}{2000}).  Some candidate NEOs may be real but
moving so fast that they are effectively unrecoverable due to timing
considerations, weather, and/or ephemeris uncertainties.

The most critical follow-up takes place within about three days of
discovery because the initial observations are usually insufficient to
even broadly identify the orbital properties of the new object. This
is because a small number of observations by a single observatory over
a short timespan do not contain enough information to determine the
geocentric distance of the object --- they contain very little
parallax information and, as a result, ephemerides rapidly become
extremely uncertain or useless.  Obtaining follow-up observations of
an asteroid is so essential that only those with confirmatory
observations are eligible for official designation and assignment of
discovery credit by the MPC.

An illustration of the need for rapid follow-up 
is asteroid 2014~JR$_{24}$ that was discovered at magnitude
$V\sim17.2$ on 2014 May 6 by the CSS. Its diameter is in the 4 to
8$\meter$ range (based on its absolute magnitude, $H=29.3$) and it
made a close approach to Earth of about 0.3~lunar distances on 2014
May 7 when it reached a peak magnitude of $V\sim15.6$. Though some astrometric follow-up was obtained there was no physical characterization
of this interesting close flyby and potential radar target, and its apparent magnitude dropped
to $V\sim27$ within three days of discovery thereby eliminating the opportunity for further follow-up. It will not have a favorable apparition for more than a decade at which time the recovery will necessarily be serendipitous because the ephemeris uncertainty will be large.

There are a variety of assets for follow-up of recently discovered
asteroids ranging from self-funded, highly-productive, sub-meter class
telescopes (\eg\ \mycitep{Birtwhistle}{2009}), to NASA-funded 1-2
meter class telescopes and, increasingly common, even access to `big
glass' in the 4 to $8\meter$ range (\mycitep{Abell}{2013}) for
characterization studies (table~\ref{tab.Top10follow-upSites}). FOVs
for the facilities extend from arc-minute scales to about a degree;
bigger fields facilitate astrometric follow-up when ephemeris
uncertainties are large.  The Astronomical Research Observatory (ARO)
located in Westfield, IL, USA, currently tops the NEO follow-up list
while at the same time contributing to astronomical education and
public outreach by involving amateur astronomers and students.  They
have recently included a southern hemisphere site at Cerro Tololo,
Chile, to follow-up objects that are too far south for most of the
follow-up facilities that are located in the northern hemisphere.
Large aperture telescopes with faint limiting magnitudes that can
gather good signal-to-noise detections even through thin clouds are
valuable tools for faint objects or when weather conditions are less
than optimal (\eg\ Mauna Kea and Magdalena Ridge in
table~\ref{tab.Top10follow-upSites}).  Unfortunately, these large
aperture observatories are typically only available for follow-up on a
few nights per lunation because they are not dedicated NEO
facilities. The researchers and sites within the University of Hawaii
network (\mycitep{Tholen \etal}{2013}, \mycitep{Wainscoat
  \etal}{2013}) give preference to follow-up of NEO candidates
discovered with the PS1 telescope using their $2\meter$ class and
larger telescopes on Mauna Kea, a big advantage for follow-up of the
smaller/fainter objects being discovered by PS1. The Magdalena Ridge
Observatory's fast-tracking $2.4\meter$ telescope (MRO; \mycitep{Ryan
  et al}{2002}) located in New Mexico, USA, has the ability to
accurately track rapidly moving targets that is essential for
follow-up of challenging faint objects or for physical
characterization (\ie\ to keep the moving object on the spectrograph's
slit). MRO performs follow-up astrometry and real-time
characterization (spin rates and composition), achieves sub-arcsec
point spread functions (PSFs) even close to the horizon (pushing the
limits to lower declinations than typical for the northern
hemisphere), and can have multiple instruments mounted simultaneously
(for rapid switching between photometry and spectroscopy).

It is common, natural, and unfortunate that a recently discovered
object receives a lot of attention primarily during the few days after
discovery when often hundreds of astrometric and physical observations
are collected by professional and amateur observers. The attention is
explicable because objects tend to be discovered near a maximum in
apparent brightness so that all aspects of follow-up are easier.
However, the major determinant in orbit element accuracy is the
temporal coverage, not the number of observations.  \ie\ there is an
opportunity cost of too many observations because most of them will
not contribute to improving the orbit.  It is therefore desirable that
observers that have access to large aperture telescopes obtain
astrometric measurements of the most important targets down to the
limiting magnitude of their instrumentation (\eg\ $V\sim26$ for 8 to
$10\meter$ class telescopes such as that achieved in the ESA's efforts
with the Very Large Telescope (VLT; \mycitep{Micheli \etal}{2014}) and Large Binocular Telescope (LBT).  Only this additional optical and orbital-arc coverage
will make it possible to determine an orbit accurate enough to recover
the object in its next apparition, that may be many years or even
decades in the future, unless radar data are available
(\eg\ \mycitep{Ostro \etal}{2004}).  The wasted effort and lack of
temporal coverage for important objects suggest that there is a need
in our community for an enhanced follow-up coordination effort beyond
the current capability of the NEOCP.  A centralized scheduling
site that assigns available observing sites to specific targets based
on weather, limiting magnitude, location, \etc, could increase the
efficiency of the overall follow-up effort (we note that this effort could be a component of the `Management Action' called out by NASA's Office of Inspector General's report on the NEO effort; IG-14-030, 2014 Sept 15).  
The ESA NEO
Coordination Centre is already coordinating follow-up observations of
high-relevance targets by a worldwide
network of cooperating observatories by providing them with rapid
triggers when an observational opportunity arises appropriate for their
systems. Over the past year their coordination efforts have contributed to the removal of 20 high-rated virtual impactors from their impact risk list.

The subsequent recovery of an object in a future apparition, that was enabled by a well-orchestrated follow-up effort in the discovery apparition, will result in a dramatic decrease in its orbit element uncertainty and correspondingly extend the ephemeris accuracy for decades into the future.  However, in some (rare) cases an object's dynamics may be so complex that, even then, predicting its motion is non-trivial.   This often occurs when there are future close encounters with major planets or when non-gravitational forces become relevant.  In these cases, observational coverage extending over multiple apparitions is often necessary to model the phenomena to a sufficient level of accuracy to accurately predict the object's future behavior (\eg\ \mycitep{Farnocchia \etal}{2013}; \mycitep{Chesley \etal}{2014};  Farnocchia \etal, this volume).

The rationale for, and scientific benefits of, observations of
asteroids at the Arecibo radar facility (Puerto Rico) and Goldstone
Deep Space Network (California, USA) are unassailable but radar
follow-up and characterization is expensive in time and
resources. Radar directly and uniquely provides range, range-rate,
shape, spin, and size data that complement the optically-derived
physical information (\eg\ \mycitep{Ostro \& Giorgini}{2004}) and
enables computation of asteroid ephemerides much further into the
future than ephemerides derived from optical-only astrometry because
of radar's exquisite precision in measuring the object's range and
range-rate.  However, the only way that a radar facility can detect an
object is with an accurate orbit provided by optical assets.  Thus,
radar is typically employed only for the most `interesting' objects.

The follow-up effort is more complicated for
space-based discoveries (\eg\ from NEOWISE; Mainzer \etal, this
volume). Delays in posting the candidate NEO's ephemerides caused by
scheduling the Deep Space Network for data transmission from the
spacecraft (by several days) can result in untenable ephemeris
errors. Furthermore, if an NEO has been discovered in a passband that
is not a ground-based standard (\eg\ an IR space telescope) then the
transformation from the spacecraft's passband to ground-based
passbands can generate significant errors in the predicted apparent
magnitudes.  The apparent magnitude can be much fainter than expected
if the object is of low-albedo (dark), such that acquiring good
signal-to-noise detections for the object is difficult or impossible
from the ground. Future wide or deep space-based surveys will need to
carefully consider ground-based follow-up requirements or build
self-follow-up into their survey strategy. 

The future needs for asteroid follow-up and rapid physical
characterization are clear --- neither will keep pace with the
expected discovery rates of future surveys (see
\S\ref{s.AnticipatingTheFuture}).  The mean $V$-magnitude of asteroids
at discovery is now about 20, which limits follow-up and
characterization sites to professional telescopes or a handful of the
most advanced amateur astronomers.  As the capabilities of the ground-
and space-based survey telescopes continue to improve, pushing the
mean discovery magnitude to even fainter values, the follow-up
contribution from the amateur community will drop.  This will require
the survey telescopes to adopt self-follow-up survey strategies that
re-acquire their own discoveries several times per month.  Indeed, the advent of deep {\it and} wide surveys has changed the cost-benefit analysis of separating surveying and follow-up such that it may now simply be more efficient to do both with the same system in an integrated discovery \& follow-up survey program.  The
re-observing strategy will reduce the possible discovery rate but the
good news is that the strategy will extend the main belt minor planet
catalog to much smaller sizes.

In many important cases it has been possible to dramatically increase the known
arc-length for an object that has become unobservable by
identifying `precovery' observations in historical images. Over the
last decade most professional observatories and surveys have developed
online repositories of all their astronomical images, including
digitizing very old photographic plates with associated astrometry and
photometry of all sources in the images.  Most of the images were
acquired for projects unrelated to asteroids but some of them contain
unrecognized detections of known asteroids.  The most obvious example
is the MPC's one-night stand file (\S\ref{s.MPC}) that is
automatically searched by the MPC for precoveries when feasible.  In
other cases, and increasingly more commonly, a survey's archival
detection database can provide precovery observations. Although the
survey's images were already inspected for moving objects there may be
cases where an object was too faint to be detected or reported with
confidence, but a precovery tracklet can be identified given a new
object's ephemeris.  The success of the precovery efforts coupled with the relatively modest cost of archiving images suggest that surveys should `save all the bits' \ie\ every bit of every pixel of every image should be stored and, even better, searchable and accessible in an online repository.

\bigskip
\refstepcounter{section}
\centerline{\textbf{\thesection. POPULATION STATISTICS}}
\bigskip

We tend to think of distinct populations of asteroids in the inner
solar system even though we know that they are actually inter-related.
For instance, the NEOs are fragments created in the collisions of main
belt asteroids that have been transported to near-Earth space by
orbital evolution driven by gravitational dynamics and thermal recoil
forces (\eg\ Binzel \etal, this volume).  Furthermore, we are now
beginning to understand that many main belt asteroids may be implanted
objects that were originally formed elsewhere in the proto-solar
system. Thus, in this section we attempt to address inner solar system
asteroids holistically, providing orbit element distributions of NEOs,
MBOs and Jupiter's Trojans, on the same scales and figures and touch
upon the transfer of objects between them.

That being said, we can not avoid the conventional nomenclature for
the populations so that when we refer to NEOs we explicitly mean those
objects with perihelion $q \le 1.3\au$ and semi-major axis $a<4.2\au$.
The MBOs are defined as those with $q > 1.3\au$ and $a<4.8\au$ while
Jovian Trojan objects (JTO) have $q>4.2\au$ and $4.8\au<a<5.4\au$.

In what follows we will frequently refer to `unbiased' and `debiased'
distributions. We use the term `unbiased' to refer to the true
distributions whereas `debiased' means that the observed distributions
have been corrected for observational selection effects. A perfect
debiasing procedure would lead to identical unbiased and debiased
distributions.  We can only provide unbiased distributions in
circumstances for which there is evidence that the entire population
is already known \eg\ main belt asteroids larger than $10\km$
diameter. A good introduction to various types of selection effects is
given by \mycitet{Jedicke \etal}{2002} and recent advances for
quantifying these effects are described by \mycitet{Jedicke
  \etal}{2014}.

We will present the asteroid's absolute-magnitude ($H$) distributions
in terms of their cumulative number, $N(<H) \propto 10^{\alpha H}$,
where $\alpha$ is referred to as the `slope'.  When available or
appropriate we may provide the diameter ($D$) number distribution
where $N(>D) \propto D^{-b}$ and $b$ is also called the `slope'.  Both
the $H$ and $D$ versions are often colloquially referred to as the
`size' distribution.  The slopes are generally not constant with $H$
or $D$ but are often approximately constant over some intervals.  The
two slope parameters are related by $b=5\alpha$ under the assumption
of a constant albedo within a population.

\bigskip
\noindent
\refstepcounter{subsection}
\textbf{ \thesubsection\ Near-Earth Objects}
\smallskip

The \mycitet{Bottke \etal}{2002} NEO population model has been
extremely successful for the past 12 years, especially considering
that it was calibrated with only 137 known NEOs. Their modeling
approach utilized dynamical constraints provided by the
orbital-element `residence-time' distributions for NEOs originating in
different source regions (essentially likelihood maps in
orbital-element space). The only notable improvement to the model was
the re-calculation of the residence-time distributions by
\mycitet{Greenstreet \etal}{2012a} using smaller time-steps and more
particles to reduce the statistical noise and provide an improved
orbital model. They did not re-fit the observed NEO population but
provided an improved NEO model using the \mycitet{Bottke \etal}{2002}
source weights and slope ($\alpha=0.35$) of the $H$ distribution. The
\mycitet{Greenstreet \etal}{2012b} NEO model suggested that there
exist dynamical pathways from prograde MBO orbits to retrograde NEO
orbits and estimated that there are about four $H<18$ NEOs on
retrograde orbits at any time. They also concluded that one of the
currently known retrograde NEOs is likely an asteroid
rather than a comet.

Preliminary results from the WISE mission suggested that there are
$981\pm18$ NEOs with $D>1\km$ and that the cumulative diameter
distribution could be represented by a broken power-law with $b=5$ for
$D>5\km$, $b=2.1$ for $1.5<D<5\km$, and $b=1.32\pm0.14$ for $D<1.5\km$
(\mycitep{Mainzer \etal}{2011b}).  Their subsequent preliminary
analysis (\mycitep{Mainzer \etal}{2012}) of the four NEO
sub-populations (Atiras, Atens, Apollos, and Amors) suggested that
there are fewer high-inclination Aten asteroids than predicted by
\mycitet{Bottke \etal}{2002}, a result that was soon verified by
\mycitet{Greenstreet \etal}{2013}.

\mycitep{Zavodny \etal}{2008} derived the absolute-magnitude and orbit
distribution for Atiras --- objects orbiting the Sun entirely interior
the orbit of the Earth (IEO) --- using data obtained by the CSS. They found
that the \mycitet{Bottke \etal}{2002} NEO orbit distribution is
consistent with the CSS observations and derived a nearly independent
measurement of the slope of the absolute magnitude distribution
($\alpha = 0.44^{+0.23}_{-0.22}$), again, consistent with both
\mycitet{Bottke \etal}{2002} and \mycitet{Stuart}{2001}.

Temporarily-captured natural Earth satellites are a
recently-recognized NEO sub-population \mycitep{Granvik
  \etal}{2012}. The average length of capture in the Earth-Moon system
is about 9 months for these `minimoons' and the largest object at any
given time has a diameter of about 1 to 2$\meter$ assuming that their
size distribution follows that observed for small Earth-impacting
asteroids (\mycitep{Brown \etal}{2002}; \mycitep{Brown \etal}{2013}).
Only one minimoon has been positively identified (2006~RH$_{120}$;
\mycitep{Kwiatkowski \etal}{2009}) due to their small sizes and rapid
sky-plane motions but the discovery rate will most likely increase in
the future as the next-generation of asteroids surveys come online
(\mycitep{Bolin \etal}{2014}, and \S\ref{s.AnticipatingTheFuture}).

Figs.~\ref{fig.Predicted-ae-ai} and \ref{fig.Known-H} illustrate the
orbit and absolute-magnitude distributions for NEOs from the
preliminary work of \mycitet{Granvik \etal}{2015} because 1) they used
the largest currently available single-survey NEO data set (from CSS)
and 2) their work is an independent data product in the sense that its
development did not rely on any other NEO models. \ie, the WISE
measurements were debiased assuming the orbital distribution by
\mycitet{Bottke \etal}{2002}, and \mycitet{Greenstreet \etal}{2012}
used the source ratios and size distributions also from
\mycitet{Bottke \etal}{2002}. The \mycitet{Granvik \etal}{2015}
estimate for the number of $H<18$ NEOs agrees with
\mycitet{Stuart}{2001}, and the number of $H<17.75$ NEOs is consistent
with the number of $D>1\km$ NEOs predicted by \mycitet{Stuart and
  Binzel}{2004} and \mycitet{Mainzer \etal}{2011}.  The advantage of
using the results of \mycitet{Granvik \etal}{2015} is that they
provide 1) the debiased $(a,e,i)$ orbital-element distributions and 2)
an extended range in the debiased absolute magnitude number
distribution to $H=25$. The functional form of their $H$ distribution
allows for a `wave' (a non-constant slope) as suggested by,
\eg\ \mycitet{Harris}{2013}, but does not require it. They also allow
a different $H$ distribution for each of their 7 source regions so
that the observable NEO $H$ distribution is the sum of 7 analytic
functions and the NEO orbit distribution changes slightly as a
function of $H$.

\bigskip
\noindent
\refstepcounter{subsection}
\textbf{ \thesubsection\ Main Belt Objects}
\smallskip

It is challenging to develop debiased size and orbit distributions for
MBOs that extend to sizes smaller than the completeness level
(currently about $H\sim17$; \mycitep{Denneau \etal}{2015}) because
asteroid families induce discontinuities into the distributions and,
contrary to NEOs, MBOs are not replenished from outside sources. That
is, the MBO population is not in a steady state --- it continues to
erode through collisional grinding, and dynamical- and
radiation-induced orbit evolution.  There are two population models
for the MBO orbit and absolute-magnitude distributions: the
Statistical Asteroid Model (SAM; \mycitep{Tedesco \etal}{2005}) and
the Pan-STARRS Synthetic Solar System Model (S3M-MBO; \mycitep{Grav
  \etal}{2011}). Both models were developed by starting from the
unbiased orbital-element distribution of large MBOs. The MBOs were
extrapolated to smaller sizes assuming a negligible correlation
between size and orbit, and that the slope of the $H$-distribution is
constant beyond the $H$ completeness limit at the time of the model's
development. The S3M-MBO uses a single slope for all sub-components of
the main asteroid belt whereas SAM employs different slopes for each
of 15 families and 3 background populations. The maximum absolute
magnitude in the S3M is correlated with perihelion distance so that it
only contains synthetic objects that can reach $V<24.5$ at perihelion
at opposition, \ie\ those that might be detectable by the 4-telescope
Pan-STARRS, equivalent to modeling all objects in the main belt with
$D\ga300\meter$ but including objects as small as $100\meter$ diameter
on the inner edge. SAM attempted to reproduce the distribution of all
MBOs with $D>1\km$.

While these models are the most comprehensive MBO models to date they
represent just the first steps in modeling this complex population.
The actual population at the smallest sizes and most extreme orbits
can not be simply extrapolated from larger objects and more common
orbits as demonstrated by:
\begin{itemize}[leftmargin=*]

\item 
SMBAS: the Sub-km Main-Belt Asteroid Survey (\mycitep{Yoshida
  \etal}{2003}) with the Subaru telescope demonstrated that the
cumulative size distribution slope is shallower for sub-km MBOs
($0.5\km<D<1\km$) than for $D>5\km$ MBOs, $b\sim1.2$ versus
$b\sim1.8$, respectively. The depletion of sub-km MBOs is more
pronounced in the outer belt ($a>2.5$au) than the inner belt
($a<2.5$au).

\item 
SKADS: the pencil-beam-type Sub-Kilometer Asteroid Diameter Survey
(\mycitep{Gladman \etal}{2009}) derived a debiased $H$-magnitude
distribution of $\alpha=0.30\pm0.02$ throughout the main belt in the
range $15 < H_R < 18$. Limiting the analysis to the inner main belt
allowed them to extend the absolute magnitude interval to $15 < H_R <
19.5$ for which the slope is marginally shallower with
$\alpha=0.23\pm0.04$. A fit to the SKADS dataset does not
\emph{require} a decrease in the slope for $H\gtrsim18$ but cannot
rule it out.

\item 
\mycitet{Terai \etal}{2013}: who suggest that high- and
low-inclination MBOs have different size-frequency distributions based
on observational data collected with the Subaru telescope. They
interpret this as a consequence of the larger impact speed for
high-inclination asteroids.

\end{itemize}

\bigskip
\noindent
\refstepcounter{subsection}
\textbf{ \thesubsection\ Trojan asteroids}
\smallskip

Trojan asteroids orbit the Sun in a 1:1 mean-motion resonance with a
planet and populate two distinct `clouds' leading or trailing the
planet by $\sim60\arcdeg$, the Lagrange 4 ($L_4$) and Lagrange 5
($L_5$) clouds, respectively. In what follows we will primarily
discuss Jupiter's Trojan objects (JTO) but also touch upon Trojans of
other planets in the inner solar system.

The Nice model (\eg\ \mycitep{Morbidelli \etal}{2005}) suggests that
the JTOs originated in the outer reaches of the protoplanetary disk
with semi-major axes in the range $15\au<a<30\au$ and were dynamically
captured by Jupiter during the repositioning of the planets after the
formation of the solar system.  Interestingly, the number of JTOs in
the clouds appears to be unequal: \mycitet{Szabo \etal}{2005}
estimated that the $L_4:L_5$ number ratio is $1.6\pm0.1$ based on SDSS
data whereas \mycitet{Grav \etal}{2011} derived an independent ratio
of $1.4\pm0.2$ using WISE measurements.  Assuming that the two values
are truly independent, the error-weighted ratio of $1.58\pm0.08$ is
more than $7-\sigma$ from unity.  The combined clouds have an
$H$-distribution with a slope of $\alpha=0.64\pm0.05$ in the range
$9<H<13.5$ (\mycitep{Szabo \etal}{2005}). The equivalent slope of the
cumulative size distribution is $b=2.2\pm0.25$ in agreement with the
preliminary result from WISE of $b\sim2$ (\mycitep{Grav
  \etal}{2011}). Similar slopes have recently also been obtained by
\mycitet{Yoshida \& Nakamura}{2005} using a substantially smaller
sample size.

Trojan objects have also been discovered for Mars (MTO; see
\mycitet{de la Fuente Marcos \& de la Fuente Marcos}{2013} and
\mycitet{Christou}{2013}, for the most recent tally), Venus (VTO;
\mycitep{de la Fuente Marcos \& de la Fuente Marcos}{2014}) and the
Earth (ETO; \mycitep{Connors \etal}{2011}). Based on the uneven
distribution of objects in the Mars $L_4$ and $L_5$ clouds (1 versus
7), and the compactness of the orbital distribution in $L_5$, both
\mycitet{de la Fuente Marcos and de la Fuente Marcos}{2013} and
\mycitet{Christou}{2013} suggest that the objects in $L_5$ have a
common origin in either a collisional or a rotational break-up
event. Whereas the 8 known MTOs are on orbits that are stable on Gyr
timescales the known ETOs and VTOs appear to be transient captures of
NEOs with kyr lifetimes.

\bigskip
\refstepcounter{section}
\centerline{\textbf{\thesection. ANTICIPATING THE FUTURE}}
\label{s.AnticipatingTheFuture}
\bigskip

The next decade of NEO discovery will likely be dominated by the large
aperture, wide FOV, ground-based telescopes or the space-based near-IR
discovery systems described below with details provided in table~\ref{tab.FutureSkySurveys}.  But this NEO-centric viewpoint belies
the important contributions to the detection and monitoring, follow-up
and characterization, of all the asteroids in the inner solar system.
Some or all of these efforts might still be dominated by smaller
aperture and visible light ground-based systems.  Furthermore, we
specifically address an emerging new camera technology that could
impact the way we detect all types of asteroids --- early tests are
promising but its actual implementation awaits funding and
verification.

This section is divided into three sub-sections that present our
current understanding of upcoming improvements, and expected new
ground- and space-based assets.  The facilities in each sub-section
are presented alphabetically.

\bigskip
\noindent
\refstepcounter{subsection}
\textbf{ \thesubsection\ Upgrades to existing facilities}

\bigskip
\noindent
\textbf{Catalina Sky Survey (CSS)}\\

The CSS team has distinguished itself in its ability to constantly
improve their systems and is currently in the process of installing
monolithic 10k$\times$10k CCD cameras on both the CSS Schmidt (703)
and MLS (G96) telescopes.  This
will effectively double the surveying capability of the CSS Schmidt
allowing 3,000 to 5,000\,deg$^2$ to be surveyed nightly to $V\sim20$.
The MLS telescope will enjoy a factor of $5\times$ increase in area
coverage through the addition of reducing optics and be capable of
surveying $\sim1,000$\,deg$^2$/night to $V\sim21.5$.  It is possible that the CSS Schmidt will be used to survey most of the observable sky over the course of a night or two.  The repeated survey coverage will yield a more complete main belt
minor planet catalog because it will be easier to link asteroid detections from night-to-night and then to other apparitions.

Further improvements will be realized by pushing the capability of
linking detections to faster rates and non-linear motion.  \ie\ most
surveys currently require that detections on a single night follow a
linear trajectory even though the closest object's paths will be
curved on the sky plane: since most surveys ignore the curvature it
implies that many nearby, perhaps even geocentric, objects are not
being identified by the current surveys.  Yet another seemingly
mundane $\sim30$\% reduction in the time required for the telescope to
step+settle+readout will yield a 10 to 12\% increase in overall area
coverage and, presumably, discovery rate.

The combined CSS and MLS telescopes should extend the completeness of
the main belt minor planet catalog to an absolute magnitude well
beyond the current $17.5$ (\mycitep{Denneau \etal}{2015}), perhaps as
deep as $H\sim19$ (about $800\meter$ diameter).  The increased
discovery rate is not expected to have any impact on the MPC
processing of the reported observations.

\bigskip
\noindent
\textbf{Panoramic Survey Telescope and Rapid Response System (Pan-STARRS)}\\

Beginning in March 2014 and continuing through at least March 2015 PS1
has been dedicated 100\% to the NEO survey after which it will
continue to devote 90\% of its time to the search.  It will be joined
in late 2014 by a second, nearly identical, telescope, \PStwo, and
both telescopes will be used for the NEO survey at the 90\% level
through at least September 2017.  Doubling the number of telescopes
will not double the NEO discovery rate because not all areas of the
sky are equally rich in unknown NEOs.  

Upgrading the CCDs in both telescopes has the potential to dramatically increase the NEO discovery rate but depends on whether funding can be identified.  The existing CCDs each have a cell structure consisting of an $8\times8$ grid of $600\times600$ pixels that make the system less efficient for finding faster moving objects because they can move over the cell gaps (which are not sensitive to light), dividing the asteroid's trail in two.  Larger monolithic commercial grade low-noise CCDs will eliminate this problem, correct cosmetic problems in the existing CCDs that decrease the effective fill factor, and allow deeper NEO searches due to lower noise.  

The image processing pipeline (IPP;
\mycitep{Magnier}{2006}) and moving object processing system
(MOPS; \mycitep{Denneau \etal}{2013}) are both being continually tuned
and improved to enhance the NEO detection efficiency.

\bigskip
\noindent
\refstepcounter{subsection}
\textbf{\thesubsection\ Exciting new facilities \& technologies}
\label{ss.NewFacilities}
\smallskip

\bigskip
\noindent
\textbf{Asteroid Terrestrial-impact Last Alert System (ATLAS)}\\

The University of Hawaii's ATLAS project (\mycitep{Tonry}{2011}) is
expected to be operational in 2015 when it will begin robotically
surveying the entire sky multiple times each night.  The system will
have two identical $50\cm$ Wright-Schmidt telescopes each located in
separate domes, one on Mauna Loa, Hawaii, USA, and the other on
Haleakala, Maui, USA.  The telescopes will have $7.4\arcdeg$ FOVs
instrumented with STA-1600 110-megapixel CCDs.  They expect to
identify asteroids roughly in the range $13<r<20$ with each primary
system and to identify asteroids in the $6<r<14$ range with auxiliary
cameras co-mounted on each of the primary telescopes (The $r$
wavelength band is similar to the Johnson-Cousins $R$ band).  Thus,
this system could monitor the entire main belt visible in the night
sky to about $r=20$ multiple times each night.

\bigskip
\noindent
\textbf{Fly-Eye}\\

The ESA is planning a network of small telescopes with a novel
`Fly-Eye' optical concept to completely scan the sky each night to
identify space debris and NEOs (\mycitep{Cibin \etal}{2012}).  The system's name captures the idea that the optics
create 16 sub-images of a field that are seamlessly stitched together
to create a composite image with a 100\% `fill-factor' (the fraction
of the image plane or sky that is actively instrumented --- \eg\ the
PS1 mosaic camera system has a fill-factor of $\sim75$\%
(\mycitep{Denneau \etal}{2013}).  A prototype system is currently
being built that has a $6.7\arcdeg\times6.7\arcdeg$ FOV (about
$45\deg^2$) with performance equivalent to a $1\meter$ diameter
telescope.

\bigskip
\noindent
\textbf{Gaia}\\

The ESA's Gaia mission (\eg\ \mycitep{Lindegren \etal}{2008}) will
measure the positions of about a billion stars spanning the entire sky
to about 24$\mu$as precision.  As of 2014 June 13 there were already
over 260 published, refereed, articles about the Gaia mission despite
the fact that the primary survey had not even begun --- 35 of those
papers had both Gaia in the title and `asteroid' as a keyword.  There
is no doubt that Gaia's astrometric and photometric catalog will
transform our ability to measure an asteroid's position and brightness
and the spacecraft should provide spectrophotometry of asteroids with
$V<20$.  Gaia is expected to detect about 350,000 asteroids and yield
spectral classifications for about a quarter of that sample
(\eg\ \mycitep{Delbo \etal}{2012}; \mycitep{Mignard \etal}{2007}).

\bigskip
\noindent
\textbf{Large Synoptic Survey Telescope (LSST)}\\

The LSST has the potential to be an extremely powerful asteroid
detection and discovery telescope.  With an expected limiting
magnitude of $R\sim24$ it could discover millions of new MBOs and
hundreds of thousands of NEOs during its operational lifetime.  The
system is expected to repeat the same fields every four nights, thus
providing its own follow-up.  A system this robust could obviate most
other ground-based observing systems for discovery (but asteroids
brighter than $R\sim18$ would saturate the system and probably not be
reported).  Its actual performance for discovering asteroids will remain to
be demonstrated because LSST intends to
acquire only 2 frames/night separated by a small time interval.  This
strategy should pose little difficulty for linking detections to known
objects but its application to discovering new objects could be
limited by the false detection rate.  

\bigskip
\noindent
\textbf{Space Surveillance Telescope (SST)}\\

The US Defense Advanced Research Projects Agency's (DARPA) SST has a
$3.5\meter$ diameter mirror and an $f/1.0$ Mersenne-Schmidt optical
system that is currently located on Atom Peak, NM, USA, at
2,400$\meter$ elevation within the White Sands Missile Range but is
scheduled to be moved to Australia.  The extremely fast telescope has
a curved focal plane instrumented with {\it curved} CCDs developed
specifically for the purpose.  It is capable of covering several
thousand deg$^2$/night to $V>20$.  The SST's primary goal is to survey
the geosynchronous belt but DARPA is considering providing some of the
survey time to an asteroid survey (\mycitep{Monet \etal}{2013}), much
like the LINEAR system (\mycitep{Stokes \etal}{2000}) was tasked for
NEO surveying when not searching for satellites.  LINEAR set the
performance bar when it joined the NEO search and there is little
doubt that the SST could do the same if it brings its new-technology,
large-aperture, large-field, fast-optics, and rapid-readout system to
bear on the NEOs.

\bigskip
\noindent
\textbf{Synthetic tracking}\\

\mycitet{Shao \etal}{2014} describe the application of two new
technologies to asteroid detection that might, if successful, advance
the field over the next decade.  They combine new low-noise
fast-readout CCDs with the computational power of graphics processing
units (GPU).  Their CHIMERA camera with a $2.5\arcmin\times2.5\arcmin$
FOV has already been tested on the Palomar $5\meter$ telescope where
they obtained images at $2\,\Hz$ (much faster rates will be used in
standard operations) and then implemented the standard shift-and-add
technique on their GPU to `synthetically track' an asteroid.  They
hope to upgrade their camera to a $8\arcmin\times8\arcmin$ FOV in the
next couple years which will enable rapid searching, albeit over
limited areas, for fast-moving asteroids to as faint as $V\sim25$.

\bigskip
\noindent
\refstepcounter{subsection}
\textbf{\thesubsection\ Possible future space-based missions}

\bigskip
\noindent
\textbf{Near-Earth Object Camera (NEOCam)}\\

NEOCam is a proposed NASA Discovery-class mission whose goal is to
survey for NEOs from a space-based platform operating from near the
Earth-Sun L1 point.  Given the success of the NEOWISE mission
(\mycitep{Mainzer \etal}{2011a}) we expect that the substantially
larger telescope surveying 100\% of the time for NEOs will be
extremely successful.  NEOCam should excel at detecting NEOs due to
their strong emission in their two proposed 6 to $10\um$ band.
Loosely translating this to $R$-band suggests that the NEOCam survey
could reach a ground-based equivalent depth of $R\sim24-25$, making it
an extraordinarily powerful survey capable of discovering hundreds of
thousands of NEOs and more than $10^6$ MBOs.  NEOCam plans to adopt a
self follow-up cadence such that each field will be imaged at least
$4\times$ per day and each field will be visited every $\sim6$ nights.
Preliminary tests of the observing cadence have already been
accomplished with the MPC and their existing software can easily link
NEOs observed in this fashion.

\bigskip
\noindent
\textbf{Sentinel}\\

The philanthropic non-profit B612 organization ({\tt
  https://b612foundation.org/}) is attempting to raise private
donations to fund the construction, launch and operations of a
spacecraft specifically designed to identify `threatening asteroids
whose orbits approach Earth', \ie\ PHOs, and identify many other NEOs
and MBOs in the process.  They hope to launch the spacecraft in
2017-2018 into a Venus-like elliptical orbit for its $6.5\yr$ mission.
The orbit is particularly effective for identifying PHOs since they
must pass through a torus of $0.05\au$ diameter centered on Earth's
orbit that is always {\it exterior} to the spacecraft's orbit.  They
expect to survey $\sim165\,$deg$^2$/hour with a 24~million pixel IR
camera ($5-10.2\um$) that will have an $11\,$deg$^2$ FOV.

\bigskip
\refstepcounter{section}
\centerline{\textbf{\thesection. SUMMARY \& CONCLUSIONS}}
\bigskip

The NEO surveys have improved dramatically since publication of
Asteroids III, yielding $8\times$ more NEOs and $10\times$ more MBOs
in the last five years than in the five years before Asteroids III.
This accomplishment is all the more impressive considering that none
of the current survey telescopes was designed with NEO observations in
mind --- they are all repurposed instruments originally designed for
entirely different operations.

The situation could change in the next decade with the introduction of
new ground-based and/or space-based surveys specifically designed for
NEO observations (\eg\ ATLAS, NEOCam, Sentinel).  These new facilities
suggest that we can expect an increased discovery
rate of asteroids in the near future but they will require a
concomitant operational evolution.  For instance, existing NEO surveys
still rely on human vetting of asteroid tracklet candidates but this
process might not be tenable if the discovery rate increases another
order of magnitude.  It seems clear that the surveys will need to
implement regular self-follow-up, improved software processing, and
 coordinate their surveying and follow-up efforts through a centralized organization ({\it cf.} NASA's Office of Inspector General's report on the NEO effort; IG-14-030, 2014 Sept 15).

Improvements to the existing surveys and the development of the next-generation surveys will push the completeness limits of all the inner solar system's asteroid sub-populations to ever smaller sizes to allow us to study the unbiased orbit and size distributions.  Debiasing the populations to even smaller sizes requires attention to the surveying plan and measuring the system's detection efficiency from the outset.  Indeed, the limiting factor in generating good debiased models of the inner solar system's asteroid populations is the lack of well-calibrated survey data \eg\ detection efficiencies, per detection astrometric and photometric uncertainties, \etc. 

Most of the largest, globally-devastating potentially hazardous
objects larger than $1\km$ diameter are now known but there remains a
residual impact risk from the remaining unknown $\sim10$\% and sporadic
long-period comets.  Perhaps surprisingly, that risk is now comparable to that of {\it sub}-$100\meter$ diameter objects because of 1) the reduced slope of the NEO SFD in the $100\meter$ to $1\km$ diameter range and 2) the partial discovery of some of the objects in the intermediate range.  In any event, there are active plans and funding to address the residual risk in the next decade, mostly funded by NASA and mostly based in
the United States.  The scientifically interesting and space-based
resource utilization options for the much more numerous objects of
$<100\meter$ diameter remains to be explored.  About a third of the
discovered NEO population is $<100\meter$ in diameter (almost 4,000
objects!) but about 3/4 of them have a MOID$<0.05\au$, placing them in
the sub-PHO category only by virtue of their mass.  These objects may
be discovered days to weeks before impact by ground-based surveys like
the upcoming ATLAS and LSST projects but the {\it maximum}
ground-based detection efficiency for objects on the scale of the
Chelyabinsk impactor is only about 50\% because about half of them
will approach Earth from the direction of the Sun --- like the
Chelyabinsk impactor.  The only way to detect and characterize these
objects before impact is with a space-based detection system like
NEOCam or Sentinel.

It is likely that physical characterization of asteroids will lag ever
farther behind their discovery simply because of the much larger
phase-space of possible `physical characterization' and the dedicated
resources that must be applied for each measurement to individual
asteroids.  It seems that the only solution is to continue the process
of limiting physical characterization to the most interesting targets
and developing new systems that allow characterization in a
multi-object mode.  For instance, NEOWISE, NEOCam and Sentinel have or
plan to measure accurate asteroid diameters in the IR while they
survey for unknown NEOs.  Similarly, the already-operational Gaia
mission will measure main belt asteroid masses, densities, shapes,
pole orientations and the impact of the Yarkovsky effect on their
orbital evolution.

In closing, the only certainty about the future of asteroid discovery,
characterization, and exploration, is that we can expect great
progress in the next decade leading to the publication of Asteroids V.

\bigskip
\textbf{ Acknowledgments.} 
\bigskip

We wish to thank E.~Christensen (CSS), L.~Denneau (Pan-STARRS),
R.~McMillan (Spacewatch), G.~Stokes (LINEAR), and R.~J.~Wainscoat
(Pan-STARRS) for providing information and perspective on their
surveys.  C.~Hergenrother and A.~W.~Harris (USA) provided helpful feedback as reviewers.



\bigskip
\centerline\textbf{ REFERENCES}
\bigskip
\parskip=0pt
{\small
\baselineskip=11pt


\refs Abell, P., and 13 co-authors (2013) The Mission Accessible Near-Earth 
Objects Survey (MANOS).\ {\em AAS/Division for Planetary Sciences Meeting 
Abstracts 45}, \#208.30.

\refs Benner, L.~A., Brozovic, M., Giorgini, J.~D., Jao, J.~S., Lee, C.~G., 
Busch, M.~W., and Slade, M.~A.\ (2013) Goldstone Radar Images of Near-Earth 
Asteroid 2012 DA14.\ {\em AAS/Division for Planetary Sciences Meeting 
Abstracts 45}, \#101.02.

\refs Birtwhistle, P.\ (2009) Lightcurves for Five Close Approach 
Asteroids.\ {\em Minor Planet Bulletin 36}, 186-187.

\refs Bolin, B., Jedicke, R., Granvik, M., Brown, P., Howell, E., Nolan, 
M.~C., Jenniskens, P., Chyba, M., Patterson, G., and Wainscoat, R.\ (2014) 
Detecting Earth's temporarily-captured natural 
satellites-Minimoons.\ {\em Icarus 241}, 280-297.

\refs Bottke, W.~F., Morbidelli, A., Jedicke, R., Petit, J.-M., Levison, 
H.~F., Michel, P., and Metcalfe, T.~S.\ (2002) Debiased Orbital and 
Absolute Magnitude Distribution of the Near-Earth Objects.\ {\em Icarus 156 
}, 399-433.

\refs Bowell, E.~and Muinonen, K.\ (1994) Earth-crossing Asteroids and 
Comets: Groundbased Search Strategies.\ {\em Hazards Due to Comets and 
Asteroids} , 149.

\refs Bowell, E., Koehn, B.~W., Howell, S.~B., Hoffman, M., and Muinonen, 
K.\ (1995) The Lowell Observatory Near-Earth-Object Search: A Progress 
Report.\ {\em Bulletin of the American Astronomical Society 27}, 1057.

\refs Brown, P., Spalding, R.~E., ReVelle, D.~O., Tagliaferri, E., and 
Worden, S.~P.\ (2002) The flux of small near-Earth objects colliding with 
the Earth.\ {\em Nature 420}, 294-296.

\refs Brown, P.~G., and 32 co-authors (2013) A 500-kiloton airburst over 
Chelyabinsk and an enhanced hazard from small impactors.\ {\em Nature 503 
}, 238-241.

\refs Chesley, S.~R., and 15 co-authors (2014) Orbit and bulk density of 
the OSIRIS-REx target Asteroid (101955) Bennu.\ {\em Icarus 235}, 5-22.

\refs Chesley, S.~R.~and Spahr, T.~B.\ (2004) Earth impactors: orbital 
characteristics and warning times.\ {\em Mitigation of Hazardous Comets and 
Asteroids}, 22.

\refs Christou, A.~A.\ (2013) Orbital clustering of martian Trojans: An 
asteroid family in the inner Solar System?.\ {\em Icarus 224}, 144-153.

\refs Cibin, L., Chiarini, M., Milani Comparetti, A., Bernardi, F., 
Ragazzoni, R., Pinna, G.~M., Zayer, I., Besso, P.~M., Rossi, A., and Villa, 
F.\ (2012) Wide Eye Debris telescope allows to catalogue objects in any 
orbital zone ..\ {\em Memorie della Societa Astronomica Italiana 
Supplementi 20}, 50.

\refs Connors, M., Wiegert, P., and Veillet, C.\ (2011) Earth's Trojan 
asteroid.\ {\em Nature 475}, 481-483.

\refs Cunningham, C.~J.\ (1988) Introduction to asteroids : the next 
frontier.\ {\em Richmond, Va., U.S.A.~: Willmann-Bell, c1988.}.

\refs Cutri, R.~M., and 41 colleagues (2012) {\em Explanatory Supplement to the 
WISE All-Sky Data Release Products , 1.

\refs de la Fuente Marcos, C.~and de la Fuente Marcos, R.\ (2014) Asteroid 
2013 ND$_{15}$: Trojan companion to Venus, PHA to the Earth.\ {\em Monthly 
Notices of the Royal Astronomical Society 439}, 2970-2977.

\refs de la Fuente Marcos, C.~and de la Fuente Marcos, R.\ (2013) Three new 
stable L$_{5}$ Mars Trojans.\ {\em Monthly Notices of the Royal 
Astronomical Society 432}, L31.

\refs Delbo', M., Gayon-Markt, J., Busso, G., Brown, A., Galluccio, L., 
Ordenovic, C., Bendjoya, P., and Tanga, P.\ (2012) Asteroid spectroscopy 
with Gaia.\ {\em Planetary and Space Science 73}, 86-94.

\refs Denneau, L., and 43 co-authors (2013) The Pan-STARRS Moving Object 
Processing System.\ {\em Publications of the Astronomical Society of the 
Pacific 125}, 357-395.

\refs Denneau, L. and 17 authors (2015) Observational Evidence for
Spin-Up as the Dominant Source of Catastrophic Disruption of Small
Main Belt Asteroids.  Accepted for publication by Icarus.

\refs Farnocchia, D., Chesley, S.~R., Chodas, P.~W., Micheli, M., Tholen, 
D.~J., Milani, A., Elliott, G.~T., and Bernardi, F.\ (2013) 
Yarkovsky-driven impact risk analysis for asteroid (99942) Apophis.\ {\em 
Icarus 224}, 192-200.

\refs Gehrels, T., McMillan, R., Frecker, J., Roland, E., Stoll, C., Doose, 
L., Shoemaker, E., Nozette, S., and Boesgaard, H.\ (1982) Progress Report 
on the Spacewatch Camera..\ {\em Bulletin of the American Astronomical 
Society 14}, 728.

\refs Gladman, B., Davis, D., Neese, C., Jedicke, R., Williams, G.,
Kavelaars, J.J., Petit, J.-M., Scholl, H., Holman, M., Warrington, B.,
Esquerdo G., and Tricarico, P.\ (2009) On the asteroid belt's orbital
and size distribution.\ {\em Icarus 202}, 104-118.

\refs Granvik, M., Vaubaillon, J., and Jedicke, R.\ (2012) The population 
of natural Earth satellites.\ {\em Icarus 218}, 262-277.

\refs Granvik, M., Morbidelli, A., Jedicke, R., Bottke, W.~F., \etal\ (2015)  
An improved Near Earth Object orbit and size distribution model.  in preparation for Icarus.

\refs Grav, T., Jedicke, R., Denneau, L., Chesley, S., Holman, M.~J., and 
Spahr, T.~B.\ (2011) The Pan-STARRS Synthetic Solar System Model: A Tool 
for Testing and Efficiency Determination of the Moving Object Processing 
System.\ {\em Publications of the Astronomical Society of the Pacific 123 
}, 423-447.

\refs Grav, T., and 16 colleagues (2011) WISE/NEOWISE Observations of the 
Jovian Trojans: Preliminary Results.\ {\em The Astrophysical Journal 742}, 
40.

\refs Grav, T., Mainzer, A.~K., Bauer, J.~M., Masiero, J.~R., and Nugent, 
C.~R.\ (2012) WISE/NEOWISE Observations of the Jovian Trojan Population: 
Taxonomy.\ {\em The Astrophysical Journal 759}, 49.

\refs Greenstreet, S.~and Gladman, B.\ (2013) High-inclination Atens are 
Indeed Rare.\ {\em The Astrophysical Journal 767}, L18.

\refs Greenstreet, S., Gladman, B., Ngo, H., Granvik, M., and Larson, S.\ 
(2012b) Production of Near-Earth Asteroids on Retrograde Orbits.\ {\em The 
Astrophysical Journal 749}, L39.

\refs Greenstreet, S., Ngo, H., and Gladman, B.\ (2012a) The orbital 
distribution of Near-Earth Objects inside Earth's orbit.\ {\em Icarus 217 
}, 355-366.

\refs Harris, A.\ (2008) What Spaceguard did.\ {\em Nature 453}, 
1178-1179.

\refs Harris, A.~W.\ (2007) An Update of the Population of NEOs and Impact 
Risk.\ {\em Bulletin of the American Astronomical Society 39}, 511. 

\refs Harris, A.~W.\ (2013) The value of enhanced NEO
surveys. IAA-PDC13–05–09 (Planetary Defence Conference, IAA, 2013).

\refs Helin, E.~F.~and Shoemaker, E.~M.\ (1979) The Palomar planet-crossing 
asteroid survey, 1973-1978.\ {\em Icarus 40}, 321-328.

\refs Hsieh, H.~H.~and Jewitt, D.\ (2006) A Population of Comets in the 
Main Asteroid Belt.\ {\em Science 312}, 561-563.

\refs Jedicke, R., Larsen, J., and Spahr, T.\ (2002) Observational 
Selection Effects in Asteroid Surveys.\ {\em Asteroids III}, 71-87.

\refs Jedicke, R., Morbidelli, A., Spahr, T., Petit, J.-M., and Bottke, 
W.~F.\ (2003) Earth and space-based NEO survey simulations: prospects for 
achieving the spaceguard goal.\ {\em Icarus 161}, 17-33.

\refs Jedicke, R., Bolin, B., Granvik, M., and Beshore, E. (2014) An
improved technique for quantifying observational selection effects in
asteroid surveys.  In revision for Icarus.

\refs Jenniskens, P., and 34 co-authors (2009) The impact and recovery of 
asteroid 2008 TC$_{3}$.\ {\em Nature 458}, 485-488.

\refs Jewitt, D.\ (2012) The Active Asteroids.\ {\em The Astronomical 
Journal 143}, 66.

\refs Jewitt, D., Weaver, H., Agarwal, J., Mutchler, M., and Drahus, M.\ 
(2010) A recent disruption of the main-belt asteroid P/2010A2.\ {\em Nature 
467}, 817-819.

\refs Kaiser, N., and 25 co-authors (2002) Pan-STARRS: A Large Synoptic 
Survey Telescope Array.\ {\em Survey and Other Telescope Technologies and 
Discoveries 4836}, 154-164.

\refs Kowalski, R.~A.~and Chesley, S.\ (2008) 2008 TC3.\ {\em 
IAUC 8990}, 1.

\refs Kowalski, R.~A. (2008) {\em MPEC 2008-T50 : 2008 TC3}.

\refs Kowalski, R.~A. (2014) {\em MPEC 2014-A02 : 2014 AA}.

\refs Kubica, J., Denneau, L., Grav, T., Heasley, J., Jedicke, R., Masiero, 
J., Milani, A., Moore, A., Tholen, D., and Wainscoat, R.~J.\ (2007) 
Efficient intra- and inter-night linking of asteroid detections using 
kd-trees.\ {\em Icarus 189}, 151-168.

\refs Kwiatkowski, T., and 14 colleagues (2009) Photometry of 2006 
RH$_{120}$: an asteroid temporary captured into a geocentric orbit.\ 
{\em Astronomy and Astrophysics 495}, 967-974.

\refs Larson, S., Brownlee, J., Hergenrother, C., and Spahr, T.\ (1998) The 
Catalina Sky Survey for NEOs.\ {\em Bulletin of the American Astronomical 
Society 30}, 1037.

\refs Lindegren, L., and 13 co-authors (2008) The Gaia mission: science, 
organization and present status.\ {\em IAU Symposium 248}, 217-223.

\refs Magnier, E.\ (2006) The Pan-STARRS PS1 Image Processing Pipeline.\ 
{\em The Advanced Maui Optical and Space Surveillance Technologies 
Conference}, .

\refs Mainzer, A., and 34 co-authors (2011a) Preliminary Results from 
NEOWISE: An Enhancement to the Wide-field Infrared Survey Explorer for 
Solar System Science.\ {\em The Astrophysical Journal 731}, 53.

\refs Mainzer, A., and 36 co-authors (2011b) NEOWISE Observations of 
Near-Earth Objects: Preliminary Results.\ {\em The Astrophysical Journal 
743}, 156.

\refs Mainzer, A., and 12 co-authors (2012) Characterizing Subpopulations 
within the near-Earth Objects with NEOWISE: Preliminary Results.\ {\em The 
Astrophysical Journal 752}, 110.

\refs Mainzer, A., and 34 co-authors (2014) Initial Performance of the 
NEOWISE Reactivation Mission.\ {\em The Astrophysical Journal 792}, 30.

\refs Masiero, J.~R., Grav, T., Mainzer, A.~K., Nugent, C.~R., Bauer, 
J.~M., Stevenson, R., and Sonnett, S.\ (2014) Main-belt Asteroids with 
WISE/NEOWISE: Near-infrared Albedos.\ {\em The Astrophysical Journal 791}, 
121.

\refs Micheli, M., Koschny, D., Drolshagen, G., Hainaut, O., and Bernardi, 
F.\ (2014) An ESA NEOCC Effort to Eliminate High Palermo Scale Virtual 
Impactors.\ {\em Earth Moon and Planets}, 12.

\refs Mignard, F., and 10 colleagues (2007) The Gaia Mission: Expected 
Applications to Asteroid Science.\ {\em Earth Moon and Planets 101}, 
97-125.

\refs Milani, A., Chesley, S.~R., and Valsecchi, G.~B.\ (2000) Asteroid 
close encounters with Earth: risk assessment.\ {\em Planetary and Space 
Science 48}, 945-954.

\refs Milani, A., Chesley, S.~R., Sansaturio, M.~E., Tommei, G., and 
Valsecchi, G.~B.\ (2005) Nonlinear impact monitoring: line of variation 
searches for impactors.\ {\em Icarus 173}, 362-384.

\refs Milani, A., and 11 co-authors (2012) Identification of known objects 
in Solar System surveys.\ {\em Icarus 220}, 114-123.

\refs Molotov, I., Elenin, L., Krugly, Y., and Ivaschenko, Y.\ (2010) ISON 
Near-Earth asteroids project.\ {\em 38th COSPAR Scientific Assembly 38}, 
688.

\refs Monet, D.~G., Axelrod, T., Blake, T., Claver, C.~F., Lupton, R., 
Pearce, E., Shah, R., and Woods, D.\ (2013) Rapid Cadence Collections with 
the Space Surveillance Telescope.\ {\em American Astronomical Society 
Meeting Abstracts \#221 221}, \#352.17.

\refs Morbidelli, A., Levison, H.~F., Tsiganis, K., and Gomes, R.\ (2005) 
Chaotic capture of Jupiter's Trojan asteroids in the early Solar System.\ 
{\em Nature 435}, 462-465.

\refs Morrison, D.\ (1992) The Spaceguard Survey: Report of the NASA 
International Near-Earth-Object Detection Workshop.\ {\em NASA STI/Recon 
Technical Report N 92}, 34245.

\refs Moskovitz, N., D. Scheeres, T. Endicott, D. Polishook,
R. Binzel, F. DeMeo, W.H. Ryan, E.V. Ryan, \etal (2014) The Near-Earth
Flyby of Asteroid 367943 (2012 DA14). Submitted to Nature.

\refs {Nesvorn{\'y}, D., Vokrouhlick{\'y}, D., and Bottke,
  W.~F.\ (2006) The Breakup of a Main-Belt Asteroid 450 Thousand Years
  Ago.\ {\em Science 312}, 1490.

\refs O'Brien, D.~P.~and Greenberg, R.\ (2005) The collisional and 
dynamical evolution of the main-belt and NEO size distributions.\ {\em 
Icarus 178}, 179-212.

\refs Ostro, S.~J.~and Giorgini, J.~D.\ (2004) The role of radar in 
predicting and preventing asteroid and comet collisions with Earth.\ {\em 
Mitigation of Hazardous Comets and Asteroids}, 38.

\refs Polishook, D.~and Ofek, E.~O.\ (2011) Asteroid Science with the 
Palomar Transient Factory Survey.\ {\em EPSC-DPS Joint Meeting 2011}, 872.

\refs Pravdo S.~H., Rabinowitz D.~L., Helin E.~F., Lawrence K.~J.,
Bambery R.~J., Clark C.~C., Groom S.~L., Shaklan S.~B., Kervin P.,
Africano J.~A., Sydney P., and Soohoo V. (1999) The Near-Earth
Tracking (NEOT) Program: An automatic system for telescope control,
wide-field imaging and object detection.  {\em Astron. J. 117},
1616-1633.

\refs Rabinowitz, D.~L.\ (1991) Detection of earth-approaching asteroids in 
near real time.\ {\em The Astronomical Journal 101}, 1518-1529.

\refs Ryan, E.~V., Ryan, W.~H., Romero, V.~D., and Magdalena Ridge 
Observatory Consortium Collaboration (2002) Magdalena Ridge Observatory 
(MRO) as a Tool for Asteroid Science.\ {\em Bulletin of the American 
Astronomical Society 34}, 898.

\refs Shao, M., Nemati, B., Zhai, C., Turyshev, S.~G., Sandhu, J., 
Hallinan, G., and Harding, L.~K.\ (2014) Finding Very Small Near-Earth 
Asteroids using Synthetic Tracking.\ {\em The Astrophysical Journal 782}, 
1.

\refs Spahr, T.~B., Hergenrother, C., and Larson, S.~M.\ (1993) High 
Ecliptic Latitude Asteroid and Comet Search.\ {\em Bulletin of the American 
Astronomical Society 25}, 1059.

\refs Spahr, T.~B., Hergenrother, C.~W., Larson, S.~M., and Campins, H.\ 
(1996) High Ecliptic Latitude Asteroid and Comet Surveying With the 
Catalina Schmidt.\ {\em Completing the Inventory of the Solar System 107}, 
115-122.

\refs Stokes, G.~H., Evans, J.~B., Viggh, H.~E.~M., Shelly, F.~C., and 
Pearce, E.~C.\ (2000) Lincoln Near-Earth Asteroid Program (LINEAR).\ {\em 
Icarus 148}, 21-28.

\refs Stokes, G.~H., Evans, J.~B., and Larson, S.~M.\ (2002) Near-Earth 
Asteroid Search Programs.\ {\em Asteroids III}, 45-54.

\refs Stuart, J.~S.\ (2001) A Near-Earth Asteroid Population Estimate from 
the LINEAR Survey.\ {\em Science 294}, 1691-1693.

\refs Stuart, J.~S.~and Binzel, R.~P.\ (2004) Bias-corrected population, 
size distribution, and impact hazard for the near-Earth objects.\ {\em 
Icarus 170}, 295-311.

\refs Szab{\'o}, G.~M., Ivezi{\'c}, {\v Z}., Juri{\'c}, M., and Lupton, R.\ 
(2007) The properties of Jovian Trojan asteroids listed in SDSS Moving 
Object Catalogue 3.\ {\em Monthly Notices of the Royal Astronomical Society 
377}, 1393-1406.

\refs Tedesco, E.~F., Cellino, A., and Zappal{\'a}, V.\ (2005) The 
Statistical Asteroid Model. I. The Main-Belt Population for Diameters 
Greater than 1 Kilometer.\ {\em The Astronomical Journal 129}, 2869-2886.

\refs Terai, T., Takahashi, J., and Itoh, Y.\ (2013) High Ecliptic Latitude 
Survey for Small Main-belt Asteroids.\ {\em The Astronomical Journal 146}, 
111.

\refs Tholen, D.~J., Micheli, M., Bauer, J., and Mainzer, A.\ (2013) 2009 
BD as a Candidate for an Asteroid Retrieval Mission.\ {\em AAS/Division for 
Planetary Sciences Meeting Abstracts 45}, \#101.08.

\refs Tich{\'a}, J., Tich{\'y}, M., and Ko{\v c}er, M.\ (2002) The Recovery 
as an Important Part of NEO Astrometric Follow-up.\ {\em Icarus 159}, 
351-357.

\refs Tonry, J.~L.\ (2011) An Early Warning System for Asteroid Impact.\ 
{\em Publications of the Astronomical Society of the Pacific 123}, 58-73. 

\refs van Houten, C.~J., van Houten-Groeneveld, I., Herget, P., and 
Gehrels, T.\ (1970) The Palomar-Leiden survey of faint minor planets.\ {\em 
Astronomy and Astrophysics Supplement Series 2}, 339.

\refs Vere{\v s}, P., Jedicke, R., Wainscoat, R., Granvik, M., Chesley, S., 
Abe, S., Denneau, L., and Grav, T.\ (2009) Detection of Earth-impacting 
asteroids with the next generation all-sky surveys.\ {\em Icarus 203}, 
472-485.

\refs Wainscoat R.~J., Veres P., Denneau L., Jedicke R., Micheli M.,
and Chastel S. (2013).  The Pan-STARRS search for Near-Earth Objects:
recent progress and future plans.  AAS-DPS meeting \#45, \#401.02.

\refs Vokrouhlick{\'y}, D., Bro{\v z}, M., Bottke, W.~F., Nesvorn{\'y}, D., 
and Morbidelli, A.\ (2006) Yarkovsky/YORP chronology of asteroid families.\ 
{\em Icarus 182}, 118-142.

\refs Vokrouhlick{\'y}, D.~and Nesvorn{\'y}, D.\ (2008) Pairs of Asteroids 
Probably of a Common Origin.\ {\em The Astronomical Journal 136}, 280-290.

\refs Wright, E.~L., and 37 colleagues (2010) The Wide-field Infrared 
Survey Explorer (WISE): Mission Description and Initial On-orbit 
Performance.\ {\em The Astronomical Journal 140}, 1868-1881.

\refs Yeomans, D.K. (2013) Near-Earth Asteroids: Finding Them Before
They Find Us.  Princeton University Press, pp. 61-73.

\refs Yoshida, F.~and Nakamura, T.\ (2005) Size Distribution of Faint 
Jovian L4 Trojan Asteroids.\ {\em The Astronomical Journal 130}, 2900-2911.

\refs Yoshida, F., Nakamura, T., Watanabe, J.-I., Kinoshita, D., Yamamoto, 
N., and Fuse, T.\ (2003) Size and Spatial Distributions of Sub-km Main-Belt 
Asteroids.\ {\em Publications of the Astronomical Society of Japan 55}, 
701-715.

\refs Zavodny, M., Jedicke, R., Beshore, E.~C., Bernardi, F., and Larson, 
S.\ (2008) The orbit and size distribution of small Solar System objects 
orbiting the Sun interior to the Earth's orbit.\ {\em Icarus 198}, 284-293.


\begin{deluxetable}{crllcccccc}
\tabletypesize{\small}
\tablecaption{
Top 10 NEO surveys (2003-2014). 
\label{tab.Top10NEOSurveys}}
\tablewidth{0pt}
\startdata
\hline
\hline
      &      &        &                       &                &	Aperture & $f$-ratio & FOV       &  pixel scale    & Obs.	\\
 Rank & NEOs & Survey & Location              & P.I.           &	  (m)    &           & (deg$^2$) & ($\arcsec$/pix) & Code	    \\
\hline
  1 & 2,910 & CSS & Mt. Lemmon, AZ, USA        & E. Christensen & 1.50     & 2.0       & 1.2       &   1.0     & G96 \\
  2 & 1,952 & CSS & Mt. Lemmon, AZ, USA        & E. Christensen & 0.68     & 1.8       & 8.2       &   2.5     & 703 \\
  3 & 1,364 & LINEAR & White Sands, NM, USA    & G. Stokes      & 1.0	     & 2.2       & 2.0       &   2.25    & 704 \\
  4 & 1,303 & Pan-STARRS1 & Haleakala, HI, USA & R. Wainscoat   & 1.8      & 4.0       & 7         &   0.26    & F51 \\
  5 &   574 & Spacewatch  & Kitt Peak, AZ, USA & R. McMillan    & 0.9      & 3.0       & 2.9       &   1.0     & 691 \\
  6 &   462 & CSS & Siding Spring, Australia   & S. Larson      & 0.5      & 3.5       & 4.2       &   1.8     & E12 \\
  7 &   166 & LONEOS & Anderson Mesa, AZ, USA  & E. Bowell      & 0.6      & 1.8       & 8         &   2.6     & 699 \\
  8 &   162 & NEOWISE & Earth polar orbit      & A. Mainzer     & 0.4      & 3.375     & 0.6 & 2.75\,/\,5.5$^\dagger$ & C51 \\
  9 &   119 & NEAT/AMOS  & Palomar, CA, USA    & E. Helin       & 1.2\,/\,1.2  & 3\,/\,2.0    & 5\,/\,2.0 & 1\,/\,1.25 & 644\,/\,608 \\
 10 &    95 & La Sagra Sky Survey & Granada, Spain & J. Nomen   & 3$\times$0.45 & 2.8   & 1.5  & 2$\times$1.5 \& 2 & J75 \\
\enddata
\vspace{-8mm} \tablecomments{The top 10 NEO surveys discovered $>96$\% of all NEOs identified during the time period from 2003 Jan 1 through 2014 Nov 18.  Columns are: Rank and number of NEOs discovered , Survey name or acronym,
  location of the survey site, name of the original or current principal
  investigator (PI), telescope aperture in meters, the $f$-ratio (focal
  length of primary mirror divided by its aperture), the FOV in square
  degrees, the image scale in arc seconds per pixel, and the site's
  IAU observatory code.}
\tnt{\dagger}{2.75$\arcsec$/pix in the 3.4, 4.6, and 12 um channels and 5.5$\arcsec$/pix in the 22 um channel.}
\end{deluxetable}

\begin{deluxetable}{lr}
\tabletypesize{\small}
\tablecaption{Minor Planet Center holdings as of 2014 Nov 6
\label{tab.MPCHoldings}}
\tablewidth{0pt}
\tablehead{
Type              &             objects}
\startdata
Numbered          &             415,688\\
Multi-opposition  &             133,306\\
1-opposition      &             115,042\\
2-night           &       $\sim$165,000\\
1-night           & $\oforder(3,000,000)$\\
\enddata
\vspace{-8mm} \tablecomments{The MPC data files contained a total of
   118,328,160 observations as of 2014 Nov 6. 
  }
\end{deluxetable}

\begin{deluxetable}{clcrc}
\tabletypesize{\small}
\tablecaption{Top 10 Asteroid Discovery Surveys (of all time) 
\label{tab.Top10DiscoverySurveys}}
\tablewidth{0pt}
\tablehead{
       &                       & Obs. &             &           \\
 Rank  & Survey or Discover    & code & Discoveries & Operations}
\startdata 
    1  & LINEAR                &  704 &  141,577  &  1997-2010\\
    2  & Spacewatch            &  691 &   85,338  &  1985-present\\
    3  & NEAT/AMOS           & 644/608 &  31,116  &  1995-2007\\
    4  & Mt. Lemmon Survey     &  G96 &   30,422  &  2004-present\\
    5  & CSS                   &  703 &   20,664  &  1998-present\\
    6  & LONEOS                &  699 &   19,856  &  1998-2012\\
    7  & Haleakala-AMOS        &  608 &    7,475  &  1995-2003\\
    8  & PLS/T-1,T-2,T-3       &  675 &    6,796  &  1960-1977\\  
    9  & E. Elst$^\dagger$     &  809 &    5,650  &  1983-2012\\    
   10  & Pan-STARRS 1          &  F51 &    4,684  &  2010-present\\
\enddata
\vspace{-4mm} \tablecomments{Rankings are from 2014 September 14.  
  Columns are: rank of the survey in terms
  of the number of asteroid discoveries credited to the survey; Survey
  name, acronym, or observer; the site's IAU observatory code; current
  number of asteroid discoveries credited to the survey; time period
  of survey operations.}
\tnt{\dagger}{Belgian professional astronomer}
\end{deluxetable}

\begin{deluxetable}{lccllccc}
\tabletypesize{\small}
\tablecaption{Top 10 Asteroid Astrometric follow-up Sites (2011 to 2014) 
\label{tab.Top10follow-upSites}}
\tablewidth{0pt}
\tablehead{
Observatory & Obs. & Location & Lead(s) & Telescope(s) & Limiting &   NEOs    &    NEOs   \\
            & Code &          &         &   aperture   & $V$ Mag. &   Total   &   $V>20$  \\
            &      &          &         &              &          & 2011-2014 & 2011-2014}
\startdata 
ARO             & H21 & IL, USA & R. Holmes              & 0.5m; 1.4m  & 22    & 3,418 & 2,112 (62\%) \\
Cerro Tololo    & 807 & Chile   & T. Linder, R. Holmes   & 0.41m       & 20-21 & 2,563 & 1,659 (65\%) \\
Spacewatch II   & 291 & AZ, USA & R. McMillan            & 0.9m; 1.8m  & 20-23 & 2,511 & 2,149 (86\%) \\
Schiaparelli    & 204 & Italy   & L. Buzzi               & 0.38m; 0.6m & 22    & 1,821 &   567 (31\%) \\
Sandlot         & H36 & KS, USA & G. Hug                 & 0.56m       & 22    & 1,040 &   437 (42\%) \\
Great Shefford  & J95 & England & P. Birtwhistle         & 0.4m        & 20-21 &   968 &   388 (40\%) \\
Mauna Kea       & 568 & HI, USA & D. Tholen, R. Wainscoat& 2.2m; 3.6m  & $>$23 &   937 &   827 (88\%) \\
Desert Moon     & 448 & NM, USA & B. Stevens, J. Stevens & 0.3m        & 20-21 &   896 &   314 (35\%) \\
Magdalena Ridge & H01 & NM, USA & W. Ryan, E. Ryan       & 2.4m        & 24    &   854 &   573 (67\%) \\
Tenagra II      & 926 & AZ, USA & M. Schwartz            & 0.8m        & 18-21 &   782 &   402 (51\%) \\
\enddata
\tablecomments{
The sites are ranked and listed by total number of NEOs observed at
each observatory from 2011 June thru 2014 June.  Columns are:
observatory name or acronym; the site's IAU observatory code; the
site's location; the name of the current lead, observer, or principal
investigator; telescope aperture in meters; limiting magnitude in the
$V$ band; total number of NEOs observed during the four years from
2011-2014 inclusive; and the number of NEOs observed at each facility
with $V>20$ (and the fraction of NEOs with $V>20$).  The last column
is provided to illustrate that some facilities excel at faint object
follow-up.}
\end{deluxetable}

\begin{deluxetable}{llcccccc}
\tabletypesize{\small}
\tablecaption{
Anticipated improved and new sky surveys. 
\label{tab.FutureSkySurveys}}
\tablewidth{0pt}
\tablehead{
Survey      & Site               & Lead(s)        &	Aperture & $f$-ratio & FOV     &  pixel scale  & IAU Obs.	\\
            &                    &    	          &	  (m)    &           & (deg$^2$) & ($\arcsec$/pix) & Code}
\startdata
ATLAS       & Haleakala, HI, USA & J. Tonry       & 0.5      & 2.0       &  30     &  1.86         & TBD \\
ATLAS       & Mauna Loa, HI, USA & J. Tonry       & 0.5      & 2.0       &  30     &  1.86         & T08 \\
\noalign{\vskip 2mm}  
CSS         & Schmidt, AZ, USA   & E. Christensen & 0.68     & 1.8       & 8.2     &   2.5         & 703 \\
CSS         & Mt. Lemmon, AZ, USA&       "        & 1.50     & 2.0       & 1.2     &   1.0         & G96 \\
\noalign{\vskip 2mm}  
Gaia        & Earth-Sun L2       & ESA, T. Prusti & $1.45\times0.5$ & n/a & 0.42   & 0.059$\times$0.18 & TBD \\
\noalign{\vskip 2mm}  
Fly-Eye     & TBD                & ESA SSA-NEO    & 1.1      & 2.0       & 45      & 1.5           & TBD \\
            &                    & D. Koschny \& G. Drolshagen & &      &         &               &     \\
\noalign{\vskip 2mm}  
LSST        & Cerro Pachon, Chile & LSSTC$^*$ & 8.4 (6.4$^\dagger$)& 1.2 & 9.6     &   0.22        & TBD \\
\noalign{\vskip 2mm}  
Pan-STARRS2 & Haleakala, HI, USA & R. Wainscoat   & 1.8      & 4.0       & 7       &   0.26        & F51 \\
\noalign{\vskip 2mm}  
Spacewatch  & Kitt Peak, AZ, USA & R. McMillan    & 0.9      & 3.0       & 2.9     &   1.0         & 691 \\
Spacewatch  & \hspace{11 mm}"    &       "        & 1.8      & 2.7       & 0.6     &   1.0         & 291 \\
\noalign{\vskip 2mm}  
SST         & Atom Peak, NM, USA &  G. Stokes     & 3.5      & 1.0       & 6       &   0.89        & G45 \\
\enddata
\vspace{-0.8cm} \tablecomments{\ Listed in alphabetical order by survey name.  Fly-Eye, Gaia, LSST, and SST are not focussed on NEO or asteroid discovery but could make major contributions to the inventory.  We provide site-specific information even for surveys that use multiple
  sites and/or telescopes.  The list includes only funded surveys as of publication of Asteroids IV.  The columns are: Survey, Site, Principal
  Investigator, telescope aperture in meters, the $f$-ratio (focal
  length of primary mirror divided by its aperture), the FOV in square
  degrees, the image scale in arc seconds per pixel and the IAU
  observatory code.  $^*$LSST is operated by the LSST Corporation.  $^\dagger$LSST will have an unusually large secondary mirror and `dual-purpose' primary mirror so we also provide the system's effective aperture.
  }
\end{deluxetable}

\clearpage


\begin{figure*}
\epsscale{2.0}
\plotone{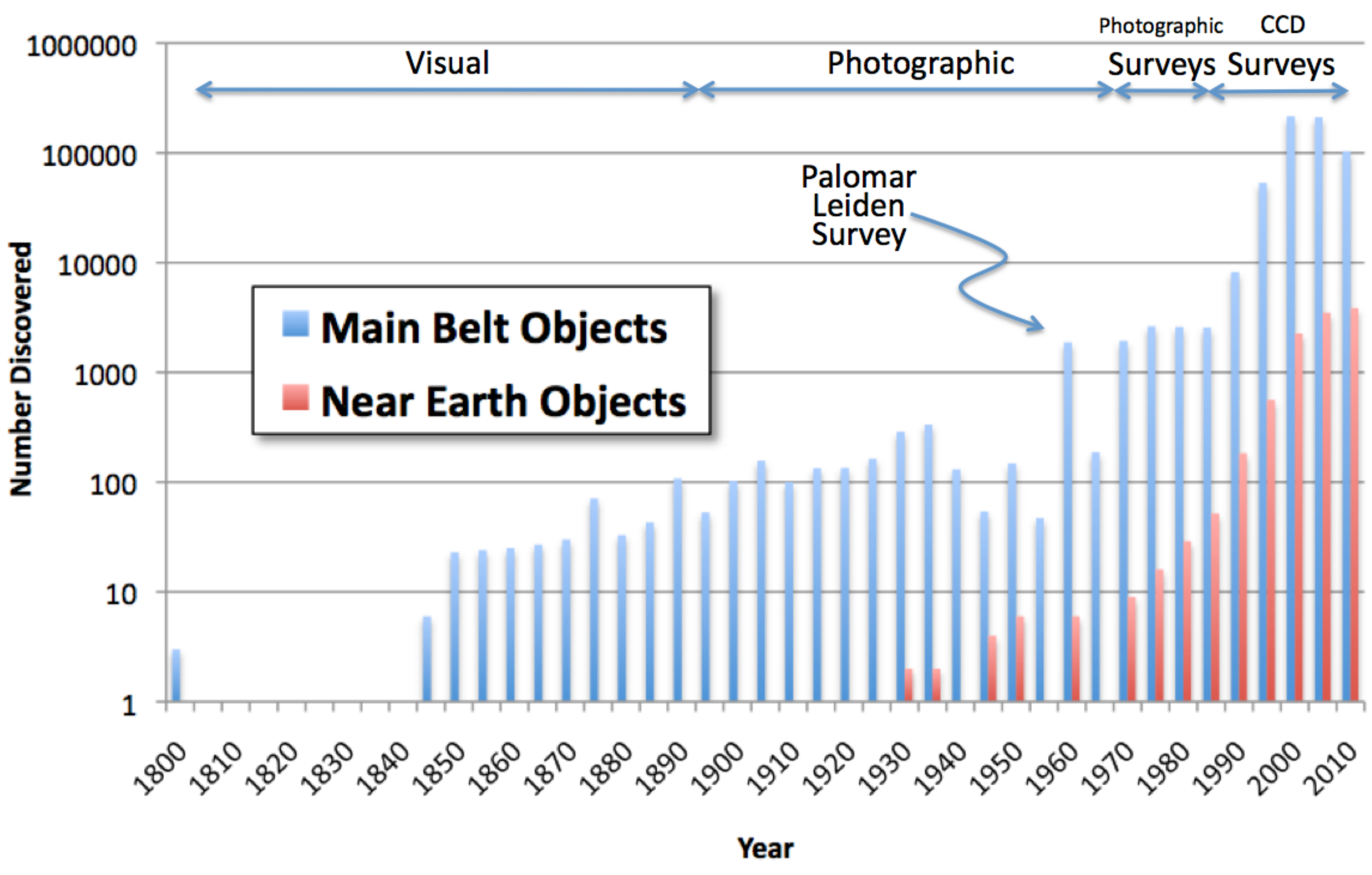}
\caption{\small 
  Main belt and near-Earth object discovery statistics
  since 1800 in five year intervals.  The time periods corresponding
  to serendipitous visual and photographic asteroid discovery is
  indicated as are the time periods for photographic and CCD surveys.
  The Palomar-Leiden Survey (PLS; \mycitep{VanHouten \etal}{1970}) in
  1960 was particularly ahead of its time as it discovered $>10\times$
  more asteroids in a few months than were being discovered over 5
  years at the time.  It took another decade before the discovery
  rates regularly matched the PLS rate. 
\label{fig.MBO+NEO-Discoveries-per-5years}}  
\end{figure*}

\begin{figure*}
\epsscale{4}
\plottwo{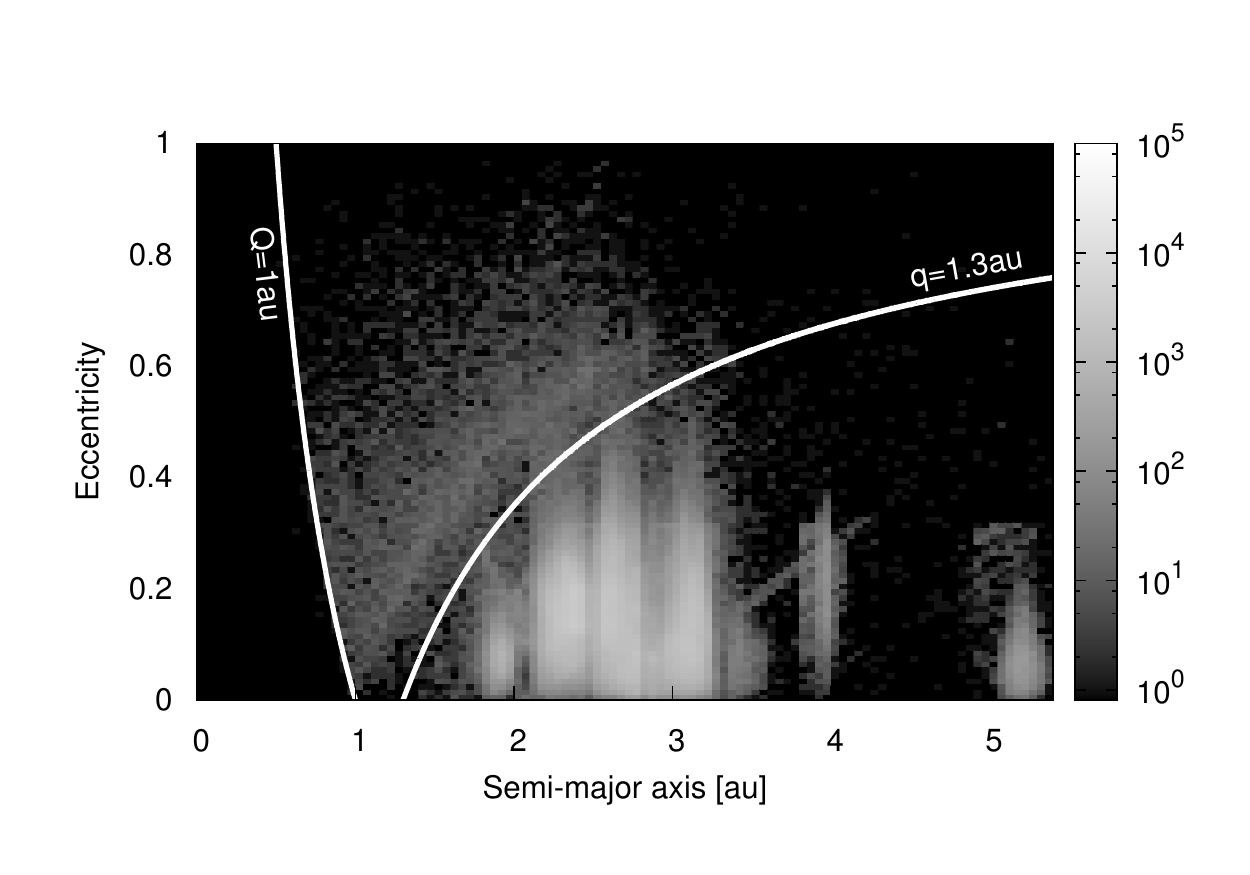}{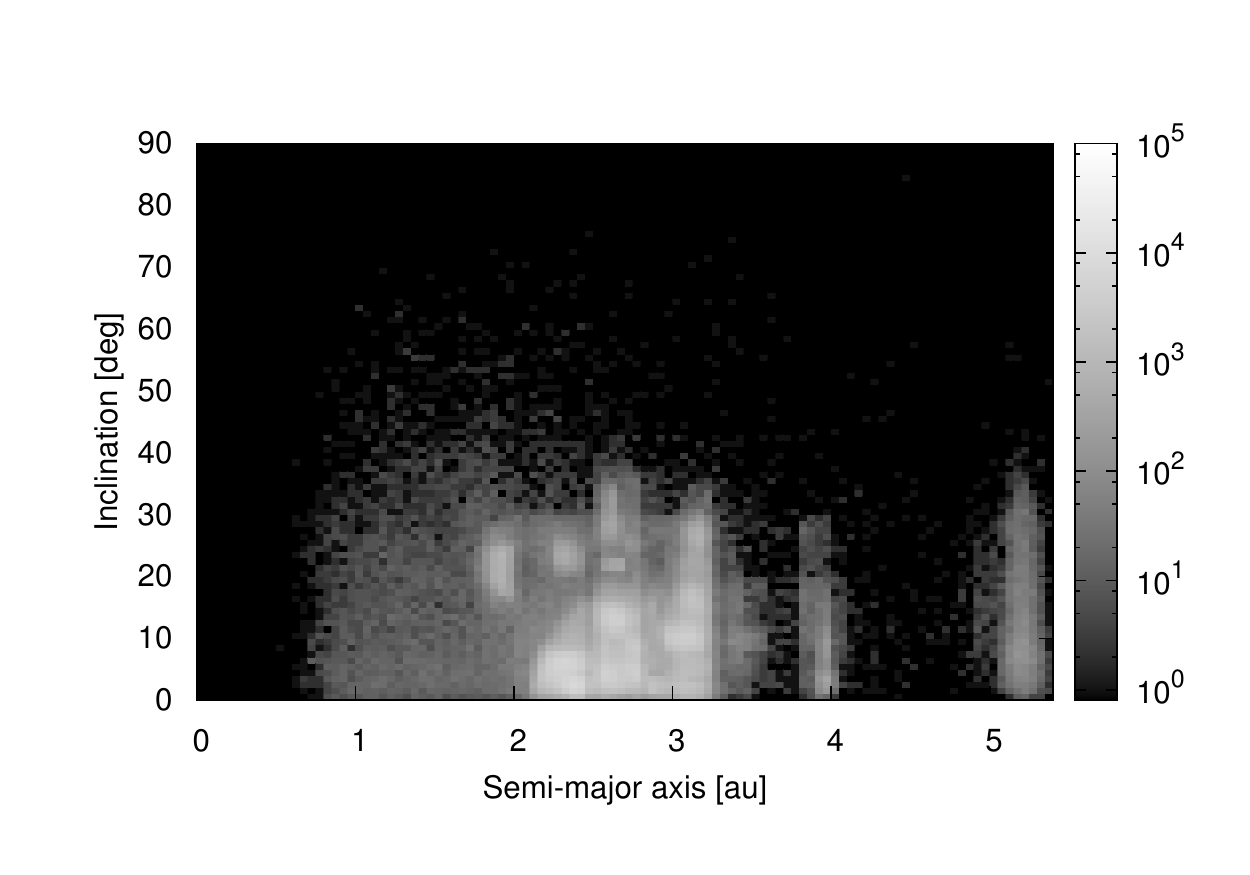}
\caption{\small (top) Eccentricity and (bottom) inclination
  vs. semi-major axis for all known asteroids in the inner solar
  system available in the MPC database as of 2014 June 8.  Note that
  the completeness, the fraction of the actual population that is
  known, varies as a function of semi-major axis so that these panels
  do not provide a good representation of the actual distribution of
  asteroids (see fig.~\ref{fig.Predicted-ae-ai}).  The clump of
  objects with $2\au\la a \la 3.5\au$ are the main belt objects and
  the clump just under $4\au$ are the dynamically separated Hilda
  population.  The group just beyond $5\au$ are the Jupiter Trojan
  objects (JTO).  The near-Earth objects are those above the (solid
  white) $1.3\au$ perihelion line in the top panel. Objects on the
  left of the aphelion line at $1\au$ in the top panel have orbits
  entirely inside the Earth's orbit. The enhancement of NEOs along the
  $q\sim1\au$ in the top panel is an observational selection effect
  that makes it easier to detect those objects from Earth.  The
  `diagonal feature' in the top panel near $a=3.7$au and $e=0.2$ are
  all WISE spacecraft discoveries that lack proper follow-up
  observations and only have a very rough, eccentricity-assumed orbit.
\label{fig.Known-ae-ai}}  
\end{figure*}

\begin{figure*}
\epsscale{4}
\plottwo{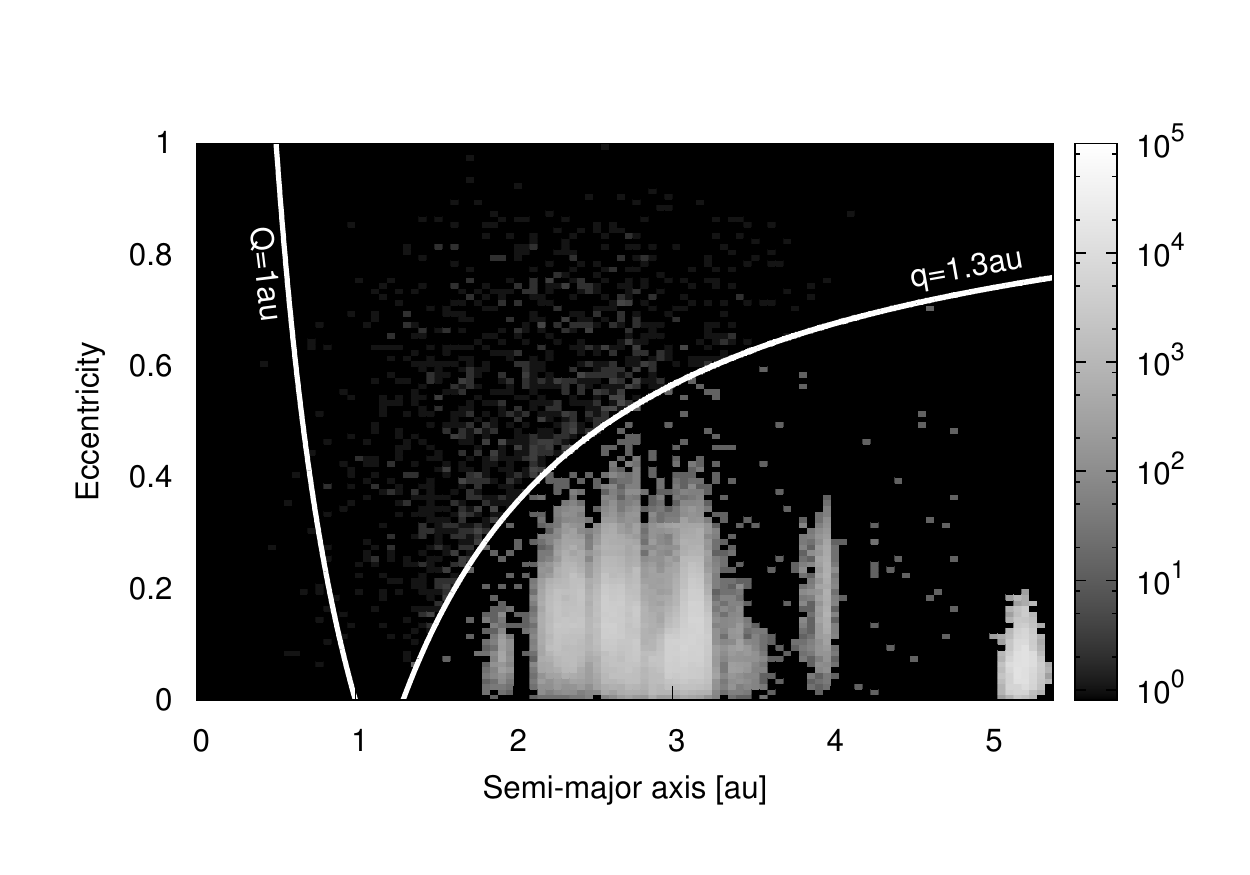}{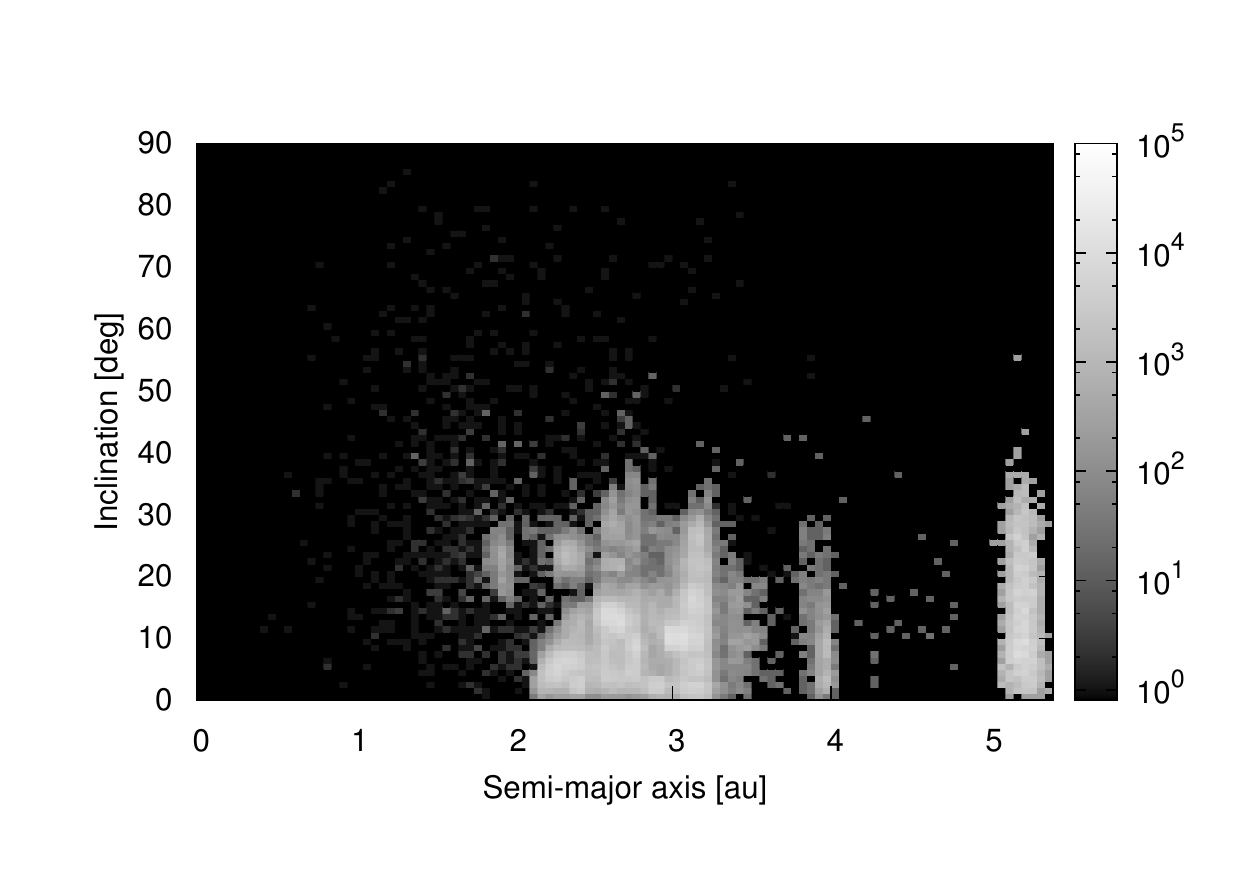}
\caption{\small Debiased (top) eccentricity and (bottom) inclination
  vs. semi-major axis for asteroids with $H<18$ in the inner solar
  system. For NEOs we use the \mycitet{Granvik \etal}{2015} model. For
  MBOs and JTOs we assume a negligible correlation between orbital
  elements and size, and extrapolate the orbit distribution from the
  assumed completeness levels of $H=15$ and $H<12.5$, respectively,
  using the slopes shown in Fig.\ \protect \ref{fig.Known-H}. The
  total number of NEOs, MBOs and JTOs with $H<18$ are predicted to be
  about 1500, 1,300,000, and 600,000, respectively. The near-Earth
  objects are those above the $1.3\au$ solid white perihelion line in
  the top panel. Objects on the left of the aphelion line at $1\au$ in
  the top panel have orbits entirely inside the Earth's orbit.
\label{fig.Predicted-ae-ai}}  
\end{figure*}

\begin{figure*}
\epsscale{2.0}
\plotone{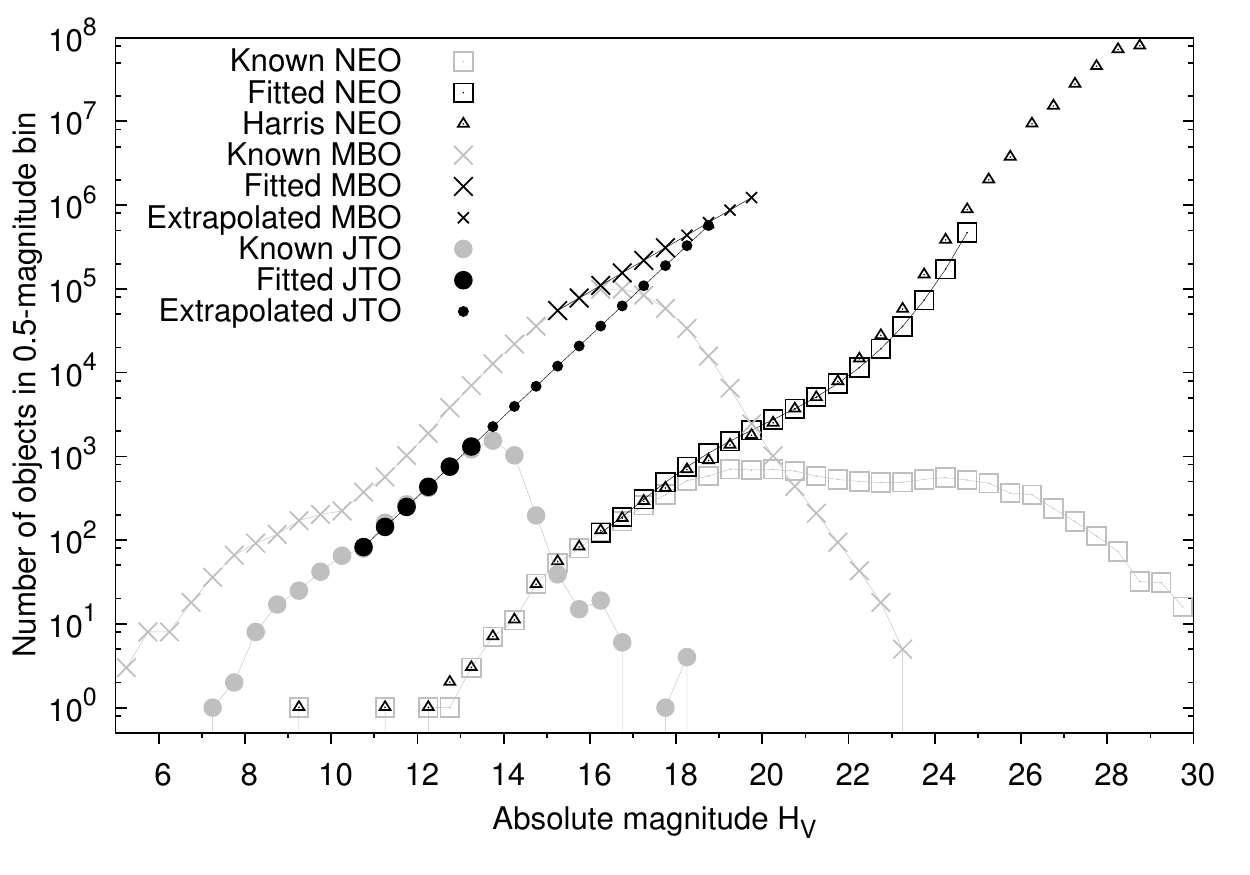}
\caption{\small 
  Known, fit, and extrapolated $H$-distributions of near-Earth (NEO),
  main belt (MBO) and Jupiter Trojan objects (JTO).  The `known'
  population is the number of objects in the MPC database as of 2014
  June 8.  The `fitted' populations represent the debiased
  distributions over specific $H$ ranges for NEOs and MBOs as derived
  by \mycitet{Granvik \etal}{2015} and \mycitet{Gladman \etal}{2009},
  respectively. The `Harris' NEO HFD is provided for reference and
  described in \mycitet{Harris \etal, this volume}. For JTOs we use a
  slope of $\alpha=0.48$ and assume that there are $10^6$ JTOs with
  $H<18.5$ --- the slope is slightly shallower but still in
  statistical agreement with that derived by \mycitet{Szabo
    \etal}{2005}. The NEO $H$-distribution has to become shallower for
  $H>25$ to match the bolide data (\mycitep{Brown \etal}{2013}). The
  `extrapolated' populations for MBOs and JTOs are simple
  linear extensions of the HFD to larger $H$ (smaller diameters)
  for the purpose of comparison with the other populations in an
  extended $H$ range.
\label{fig.Known-H}} 
\end{figure*}

\end{document}